\documentclass[manuscript,screen,nonacm]{acmart}
\usepackage{amsmath}
\usepackage{colortbl}

\usepackage{tabularray}
\usepackage{longtable}
\newcolumntype{L}[1]{>{\raggedright\arraybackslash}m{#1}}
\usepackage{geometry}
\usepackage{array}
\usepackage{multirow}
\usepackage[most]{tcolorbox}
\tcbuselibrary{breakable}
\usepackage{cite}
\usepackage{graphicx}
\usepackage{soul}
\usepackage{xcolor} 
\usepackage{xurl}
\usepackage{subfig}
\usepackage{afterpage}  %
\usepackage{hyperref} 

\usepackage{longtable}
\usepackage{makecell}
\AtBeginDocument{%
  \providecommand\BibTeX{{%
    \normalfont B\kern-0.5em{\scshape i\kern-0.25em b}\kern-0.8em\TeX}}}
\hypersetup{
    colorlinks=true, 
    urlcolor=blue,   
    linkcolor=red,   
    citecolor=green  
}
\begin{document}

\title{Systematic Literature Review of AI-enabled Spectrum Management in 6G and Future Networks}

\author{Bushra Sabir}
\author{Shuiqiao Yang}
\author{David Nguyen}
\author{Nan Wu}
\author{Alsharif Abuadbba}
\author{Hajime Suzuki}
\author{Shangqi Lai}
\author{Wei Ni}
\author{Ding Ming}
\author{Surya Nepal} 
\affiliation{
\institution{CSIRO's Data61} 
\country{Australia}
}
\email{firstname.lastname@data61.csiro.au}

\



\renewcommand{\shortauthors}{Bushra et al.}

\begin{abstract}
 Artificial Intelligence (AI) has made significant advancements across diverse domains, such as healthcare, finance, and cybersecurity, with notable successes including DeepMind's breakthroughs in medical imaging and Tesla's AI-driven autonomous vehicles. 
 As telecommunications transition from 5G towards 6G era, integrating AI becomes crucial for addressing complex demands like data processing, network optimization, and security. 
 Despite ongoing research into AI's application in next-generation networks, there is a notable research gap in consolidating advancements in AI-enabled Spectrum Management (AISM) systems.
Spectrum management (SM) is the process of optimizing the use of radio frequencies to prevent interference, improve network performance and reduce latency. However, traditional SM methods are insufficient for 6G due to dynamic, complex and network demands of 6G, necessitating AI's role in enabling spectrum optimization, security, and network efficiency that are fundamental to driving the technological leap forward.
 This study aims to address this gap by:
(i) Conducting a systematic review of AISM methodologies, focusing on learning models, data handling techniques, and performance metrics.
(ii) Examining security and privacy concerns related to AI and traditional network threats within AISM contexts.
Using the Systematic Literature Review (SLR) methodology, we meticulously analyzed 110 primary studies to:
(a) Identify utility of AI in spectrum management applications.
(b) Develop a taxonomy of AI approaches used in these applications.
(c) Classify the datasets and performance metrics utilized.
(d) Detail security and privacy threats and their countermeasures.
Our findings reveal challenges such as under-explored AI usage in critical AISM systems, computational resource demands, transparency issues, the need for real-world datasets, imbalances in security and privacy research, and the absence of testbeds, benchmarks, and security analysis tools. Addressing these challenges is vital for maximizing AI's potential in advancing 6G technology.
\end{abstract}



\maketitle

\section{Introduction}

The convergence of artificial intelligence (AI) with communication technologies is on the horizon, indicating the impending arrival of sixth-generation (6G) mobile networks. This technological advancement is poised to redefine both digital and physical experiences \citep{husen2022survey,shen2023five}. The International Telecommunication Union (ITU) aims to finalize 6G standardization by 2030, introducing AI-driven technology to create self-learning, intelligent networks that integrate the digital and physical worlds \citep{itu2020spectrum, wang2023road}.  6G seeks to revolutionize global communication by delivering high data rates, extensive connectivity, improved cost-efficiency, efficient resource management, and enhanced security, primarily through the application of AI and other cutting-edge technologies \citep{sheth2020taxonomy, zhang20196g}.

However, realizing 6G's potential requires expanding network capacity, which involves identifying the appropriate telecommunications spectrum for its implementation.
The term "spectrum" refers to the range of electromagnetic radio frequencies used for wireless data transmission, divided into different frequency bands for specific purposes like audio broadcasting, mobile communications, Wi-Fi, and television transmission. Although not finalized, sub-6GHz (2.4/3.5/5 GHz), mmWave (28/39/60 GHz), THz (above 100 GHz), and Non-Radio Frequency (RF) (laser-based optical communications, visible light communications (VLC), and quantum communications spectrum) are under consideration for 6G \citep{acma2023five,wang2023road}. Nevertheless, the spectrum is a finite and valuable resource. Its effective management is essential to ensure seamless communication \citep{stine2004spectrum, wang2023road}.

Spectrum Management (SM) is a critical aspect of modern wireless communication infrastructure, involving the careful allocation, regulation, and coordination of the radio frequency spectrum \citep{stine2004spectrum}.
Historically, wireless spectrum management relied on a complex regulatory framework and diverse policies. Nevertheless, this approach became increasingly problematic, driven by a manual process for assessing spectrum needs. However, the methods for spectrum allocation, often guided by small studies with inherent biases, resulted in sub-optimal and inflexible policies, hindering the efficient utilization of wireless spectrum \citep{matinmikko2020spectrum}.
Despite the historical challenges, the evolution of spectrum management across technological epochs is evident. In the 1G to 3G era, the focus was on analogue voice transmission, leading to a static approach in spectrum allocation. The advent of 4G marked a shift towards digital transmission and data services, necessitating more efficient spectrum utilization. Nonetheless, as we approach the 6G era, expanding network capacity poses challenges in spectrum management. While different frequency bands and innovative technologies are considered, effective spectrum management remains crucial for ensuring seamless communication.

In this evolving landscape, the application of AI into spectrum management emerges as a transformative catalyst. AI addresses challenges associated with ambitious goals of 6G, making it an essential component in the ongoing evolution of spectrum management \citep{almazrouei2020can,acma2023five}.

\subsection{Motivation for the Survey}
We are now in the era of AI, driven by three key factors: the abundance of big data, the development of advanced deep learning algorithms, and the increased computing power facilitated by graphics processing units (GPUs) \citep{lin2023embracing}. To this end, 
AI has become a powerful technology for efficient and effective spectrum management in beyond fifth-generation (5G) mobile communication systems and is anticipated to be used in 6G and future generation networks \citep{du2020machine, sagduyu2020wireless, wang2023road, sheth2020taxonomy}. 
AI-enabled Spectrum Management (AISM) approaches are pivotal in optimizing the finite radio frequency spectrum, facilitating dynamic sensing, sharing, and utilization. They adapt to ever-changing environmental conditions and user needs, enable real-time spectrum hand-offs and efficient resource allocation, enhance spectral efficiency, predict future usage patterns, and bolster wireless communication security through automated threat detection and mitigation \citep{acma2023five,ranaweera2021mec}.

The motivation of this survey is to thoroughly understand the landscape of AISM approaches from multiple perspectives. The distinct objectives for this comprehensive survey is as follows:
\begin{itemize}
    \item To systematically identify and analyze existing AI-enabled Spectrum Management systems and categorize their learning approaches, datasets, and performance measures.
    \item To investigate the security and privacy use-cases of AISM systems, considering both inherited aspects from traditional Spectrum Management approaches and novel adversarial challenges specific to AISM systems.
\end{itemize}

\subsection{Contribution of This Survey}
In this survey, we adopt a Systematic Literature Review research methodology based on the general guideline proposed by Kitchenham et al. \citep{slrguidelines} to gain a comprehensive understanding of AISM techniques in the context of beyond 5G (B5G) and 6G networks (refer to Section \ref{section-research-methodology}). We examined 110 high quality (H-index $>60$ \citep{RN71}) articles from 2013 to 2023. The main contribution of our survey is listed below:

\begin{enumerate}
    \item We provide a taxonomy and analysis of AISM applications, datasets, learning approaches and performance measures commonly utilized for training and evaluating AISM tasks. The insight reveals the strengths, weaknesses, and practical implications of these systems for potential adoption in 6G and future networks.
    \item We present a classification of the security and privacy challenges and their mitigation strategies that stem from the implementation of AISM and are inherited from the infrastructure. 
    \item We identify the challenges and propose future directions to stimulate future research.
    
\end{enumerate}
 
This SLR's insights reveal the strengths, weaknesses, and practical implications of AISM systems for potential adoption in 6G and future networks.
It offers strategic insights to aid decision-making for researchers, practitioners, and policymakers interested in spectrum management for advanced communication systems. Moreover, the survey contributes to a comprehensive understanding of the security landscape in this domain. Finally, it identifies gaps, challenges, and potential improvements to guide researchers in addressing critical aspects in the development and deployment of AISM solutions.

\subsection{Comparison to Related Surveys}
In Table \ref{comparison}, we present a comparative analysis of our SLR and previous studies.
Numerous recent surveys have extensively examined the role of AI in B5G and future-generation networks. However, the majority of these studies \citep{husen2022survey,wang2023road,sheth2020taxonomy, zhang20196g,shen2023five} primarily focused on the broader application of AI in 6G wireless communication, there is a noticeable gap in the literature specifically addressing AI's role in spectrum management. 
Notably, many surveys only addressed the overall design and architecture of AI-enabled 6G systems, as evidenced by works such as \citep{
,du2020machine, sheth2020taxonomy}.
For example, in the review article by Sheth et al. 
\citep{sheth2020taxonomy} provides a thorough survey conducted on the application of AI in 6G communication technology, showcasing its potential across various future applications. The survey covers some aspects of Dynamic Spectrum Allocation, Spectrum Sensing and RF Configuration. However, it's not focused on AI for Spectrum Management in 6G.  Similarly, \citep{du2020machine} provides a visionary architecture of machine learning approaches in the context of 6G wireless networks and explores the challenges posed by the complex architecture of 6G networks and how machine learning can help with channel estimation, spectrum management,  network orchestration and management. Additionally, the paper highlights the importance of data privacy and security in the implementation of machine learning algorithms in 6G networks. 

Conversely, certain survey papers \citep{liu2021adversarial, adesina2022adversarial,wang2023adversarial} made an effort to delve into attacks on AI-enabled wireless and mobile computing approaches but failed to encompass all aspects of AI-enabled spectrum management. For example, the survey
\citep{adesina2022adversarial} thoroughly examines Adversarial Machine Learning (AML) in wireless communications, providing a taxonomy of AML attacks. The survey covers methods for generating adversarial examples, diverse attack mechanisms, and defensive measures in wireless communication. It discusses contemporary research trends, anticipating the future landscape of AML in next-gen wireless communications. However, it only focuses on spectrum sensing and resource allocation adversarial attacks in the realm of spectrum management. 
In contrast, \citep{liu2021adversarial} systematically reviews AML applied to wireless and mobile systems, covering approaches from the physical to the application layer. Emphasizing the impact on wireless challenges, it doesn't specifically focus on spectrum management.

On the other hand, the review \citep{wang2023adversarial} offers a comprehensive overview of recent advancements in adversarial attacks and defenses within DNN-based classification models in communication applications. However, it does not address spectrum management or 6G. The survey reviews over 220 papers from 2021 to 2023, categorizing recent adversarial attack methods and defense techniques. While focused on adversarial attacks, it only touches on spectrum sensing, monitoring, and channel fading.
Nevertheless, few surveys covered spectrum management aspects for specific technologies such as UAV \citep{jasim2021survey} and IoT \citep{afzal2018unlocking}. For instance, the survey \citep{jasim2021survey} aims to fill a gap in the existing literature by providing a detailed examination of spectrum management tailored to the unique characteristics of UAV networks. It covers regulatory aspects, deployment scenarios, management tools, and potential solutions, ultimately contributing to the understanding and development of efficient and safe UAV communication systems. It does not focus on AI approaches and overall spectrum management in 5G.

Lastly, some surveys focused on specific applications of spectrum management such as resource management \citep{bartsiokas2022ml}, Spectrum sensing \citep{syed2023deep} and Dynamic Spectrum Sharing \citep{jeon2019coordinated}.
Nevertheless, the existing literature lacks a comprehensive analysis of the AI approaches in B5G, 6G and future-generation networks. Consequently, compared to the previous studies, our present work covers all proposed aspects of the AISM as summarized in Table ~\ref{comparison}.
\begin{table}[!tb]
\centering
\caption{Survey Metadata on AI-enabled Spectrum Management (AISM) Approaches [Deep Learning (DL), Adversarial Machine Learning (AML), Visionary (V), Partial (P), Not Applicable (N/A)]}
\label{comparison}
\footnotesize
\begin{tabular}{|m{12mm}|m{12mm}|m{7mm}|m{7mm}|m{7mm}|m{7mm}|m{7mm}|m{7mm}|m{7mm}|m{7mm}|m{7mm}|m{7mm}|m{6mm}|}
\hline
&\textbf{Survey} &\textbf{ Mao et al. \citep{mao2018deep}}  &\textbf{Zhang et al.\citep{zhang20196g}} & \textbf{Yang et al. \citep{yang2020artificial}} & \textbf{Du et al. \citep{du2020machine} }& \textbf{Sheth et al. \citep{sheth2020taxonomy}} & \textbf{Liu et al.\citep{liu2021adversarial} }& \textbf{Jasim et al. \citep{jasim2021survey}} & \textbf{Husen et al. \citep{husen2022survey} }& \textbf{Adesina et al. \citep{adesina2022adversarial}} & \textbf{Shen et al. \citep{shen2023five} } & \textbf{OUR} \\ 
\cline{2-13}
&{Year} &2018 & 2019 & 2020 & 2020 & 2020 & 2021 & 2021 & 2022 & 2022 & 2023  & 2024 \\ \cline{2-13}

\textbf{Meta-Data} &{Systematic} & \texttimes & \texttimes & \texttimes & \texttimes & \texttimes & \texttimes & \texttimes & \checkmark & \texttimes & \texttimes  &  \checkmark \\
\cline{2-13}
& {B5G} & \texttimes&\checkmark & \checkmark & \checkmark & \checkmark & \texttimes &\texttimes & \checkmark& \checkmark & \checkmark  &\checkmark \\ 
\cline{2-13}
&{Focus}& DL &V&V&V&V&AML&UAV&V&AML&V& AISM, AML\\ \cline{2-13}

&{PHY/MAC Layer} & \checkmark & \texttimes & \texttimes & \checkmark & \checkmark & \checkmark & \checkmark& \texttimes & \checkmark & P &  \checkmark \\ \hline

\textbf{Spectrum Mgt (SM)} &{Beam Mgt} & P & \texttimes & \texttimes & \texttimes & P & P & \checkmark & \texttimes & \texttimes& \checkmark &  \checkmark \\
\cline{2-13}

&{Channel Mgt} & P & \texttimes & \texttimes & \texttimes & P & P & \checkmark& \texttimes & P & \texttimes & \checkmark \\ \cline{2-13}

&{Resource Mgt} & P &\texttimes & \texttimes & \texttimes & \texttimes & P & \checkmark & \texttimes& P& P &  \checkmark \\  
\hline
\textbf{AI Pipeline} &{Dataset} & \checkmark &\texttimes & \texttimes & \texttimes & \texttimes & \texttimes &\texttimes & \texttimes & \texttimes & \texttimes  & \checkmark \\ \cline{2-13}

&{Learning Approaches} & P & \texttimes & \checkmark & \checkmark & \checkmark & \texttimes &\texttimes &\checkmark  & \texttimes & \checkmark& \checkmark \\ \cline{2-13}

&{Performance Measures} & \texttimes & \texttimes & \texttimes & \texttimes & \texttimes & \texttimes & \texttimes& \texttimes & \texttimes & \texttimes & \checkmark \\ \hline 
\textbf{Security and Privacy} &{Inherited} & \texttimes& \texttimes & \texttimes & \texttimes & \texttimes & \texttimes &\checkmark & \texttimes& \texttimes & P& \checkmark \\ \cline{2-13}

&{Adversarial} & \texttimes & \texttimes & \texttimes & \texttimes & \texttimes & \checkmark & \texttimes & \texttimes & \checkmark & \texttimes& \checkmark \\ \hline

\end{tabular}
\end{table}

\begin{figure*}[tb!]
\centerline{\includegraphics[width=\linewidth]{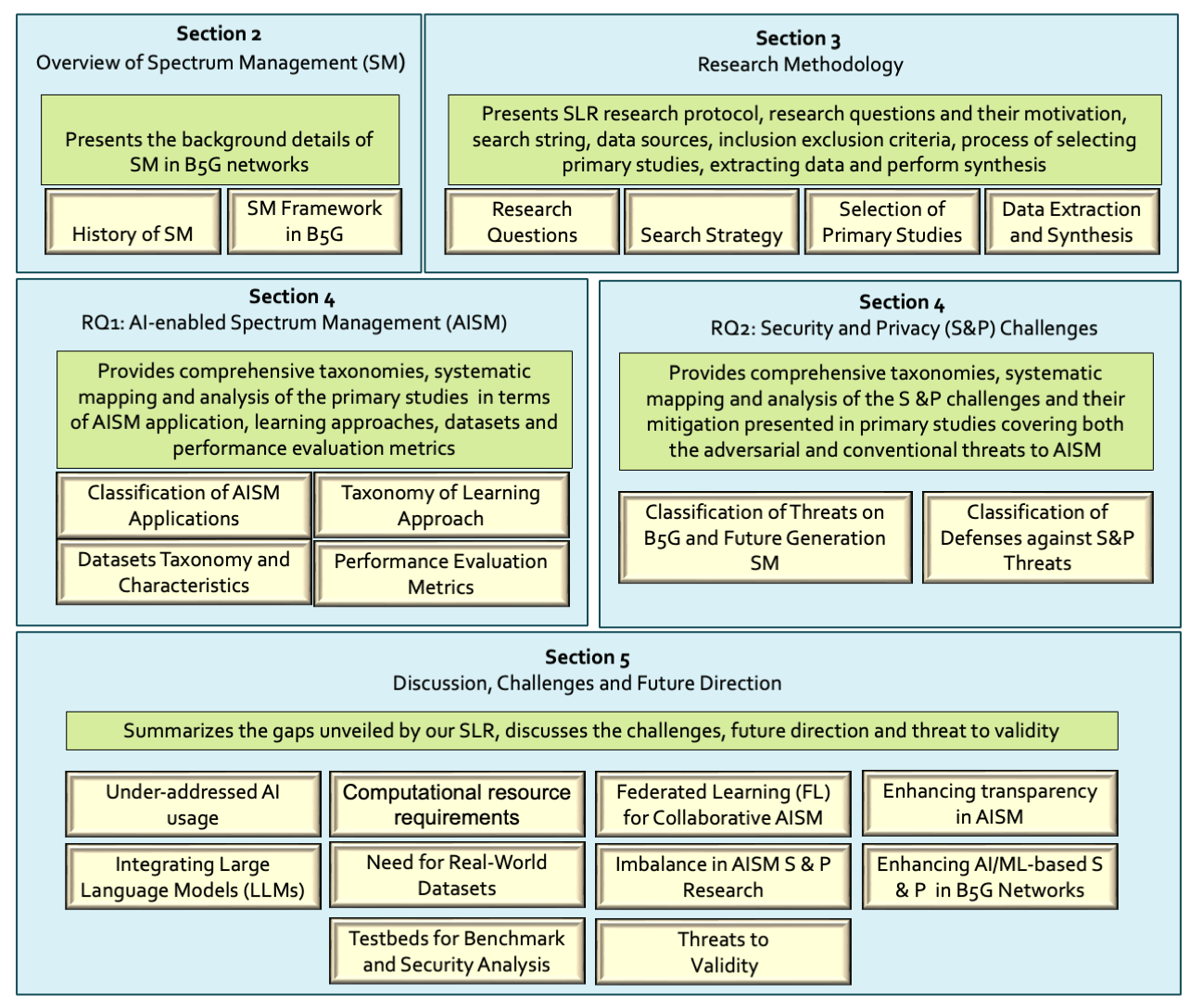}}
\caption{Diagramatic view of the organization of this SLR.}
\label{organization}
\end{figure*}

\begin{figure*}[tb!]
\centerline{\includegraphics[width=\textwidth]{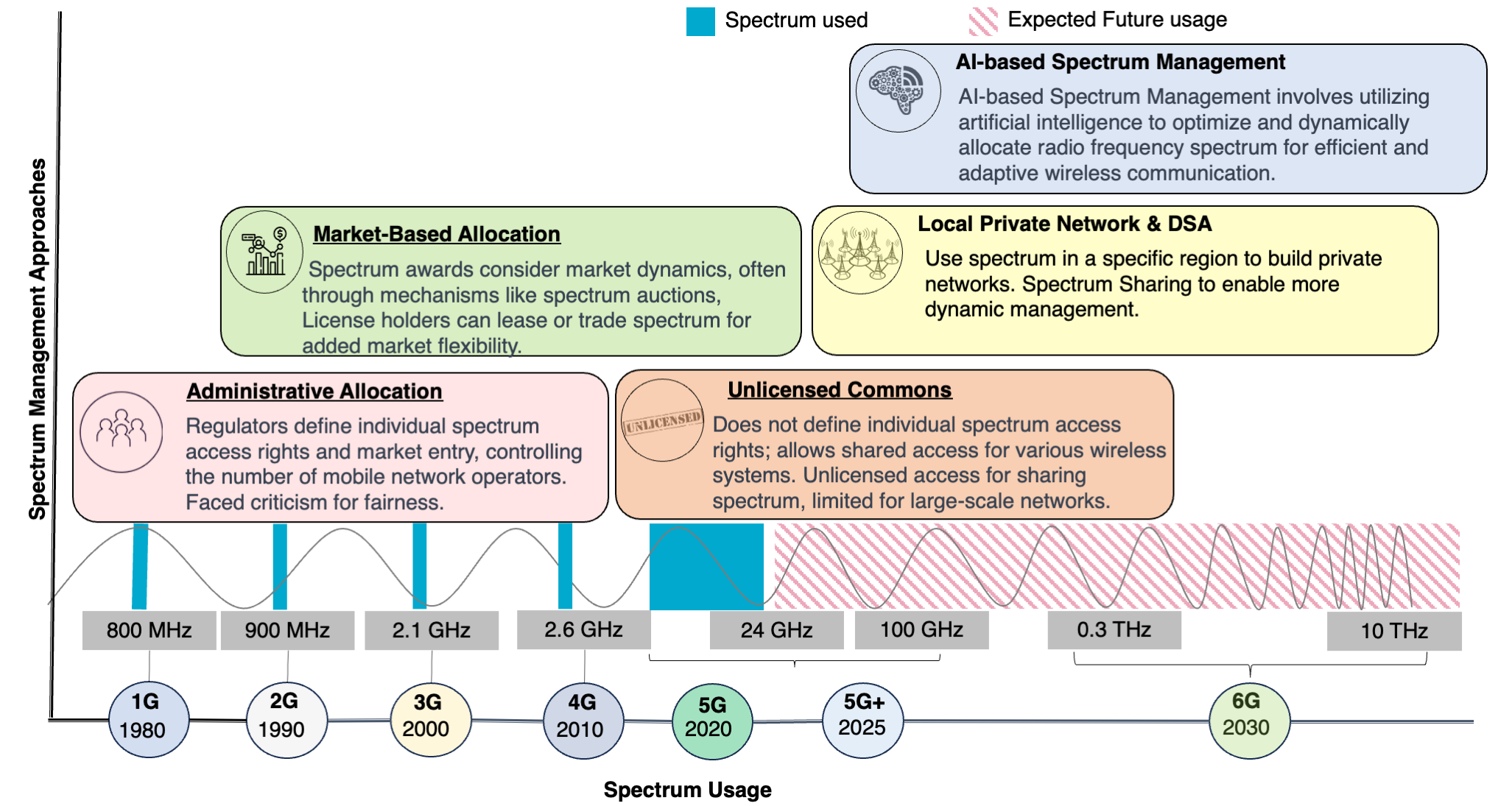}}
\caption{Evolution of Spectrum and its Management Approaches}
\label{evolve}
\end{figure*}

\section{An Overview on Spectrum Management}
In this section, we present an overview of spectrum management techniques.
We start with a brief introduction to the history of spectrum management to demonstrate how this technique evolved to fit emerging telecommunication techniques and standards.
We will then articulate the key components of state-of-the-art spectrum management, how these components enable effective spectrum management in modern telecommunication systems, and their potential issues under the context.

\subsection{History of Spectrum Management}\label{subsec:history}
The evolution of spectrum management, depicted in Figure ~\ref{evolve}, spans from the early 20th century to the current 5G era, with a trajectory toward the anticipated 6G era. In the early generations (1G and 2G), spectrum management relied on Administrative Allocation within legislative and regulatory frameworks, ensuring effective use. The 3G and 4G era saw the introduction of Market-Based Mechanisms to boost competition and economic goals, alongside the emergence of the Unlicensed Commons Approach in 4G. This supported short-range technologies like Bluetooth and Wi-Fi, driving a shift towards more accessible wireless communication and enhancing connectivity in modern digital ecosystems. The recent 5G era witnessed advancements in high-altitude platforms and satellites, prompting new spectrum allocations. Concepts such as Local and Private Networks (networks customized for specific organizational needs with enhanced connectivity, security, and services)  and Spectrum Sharing (where multiple users dynamically share spectrum resources) addressed diverse spectrum demands. Looking ahead to 6G, the landscape anticipates continued Spectrum Sharing, Unlicensed Commons use, and the integration of AI-based spectrum management (AISM) for optimized solutions and fine-grained control. The upcoming sections delve into recent AISM approaches and standardization efforts and address security and privacy challenges in this evolving domain.

\begin{figure*}[tb!]
\centerline{\includegraphics[width=\textwidth]{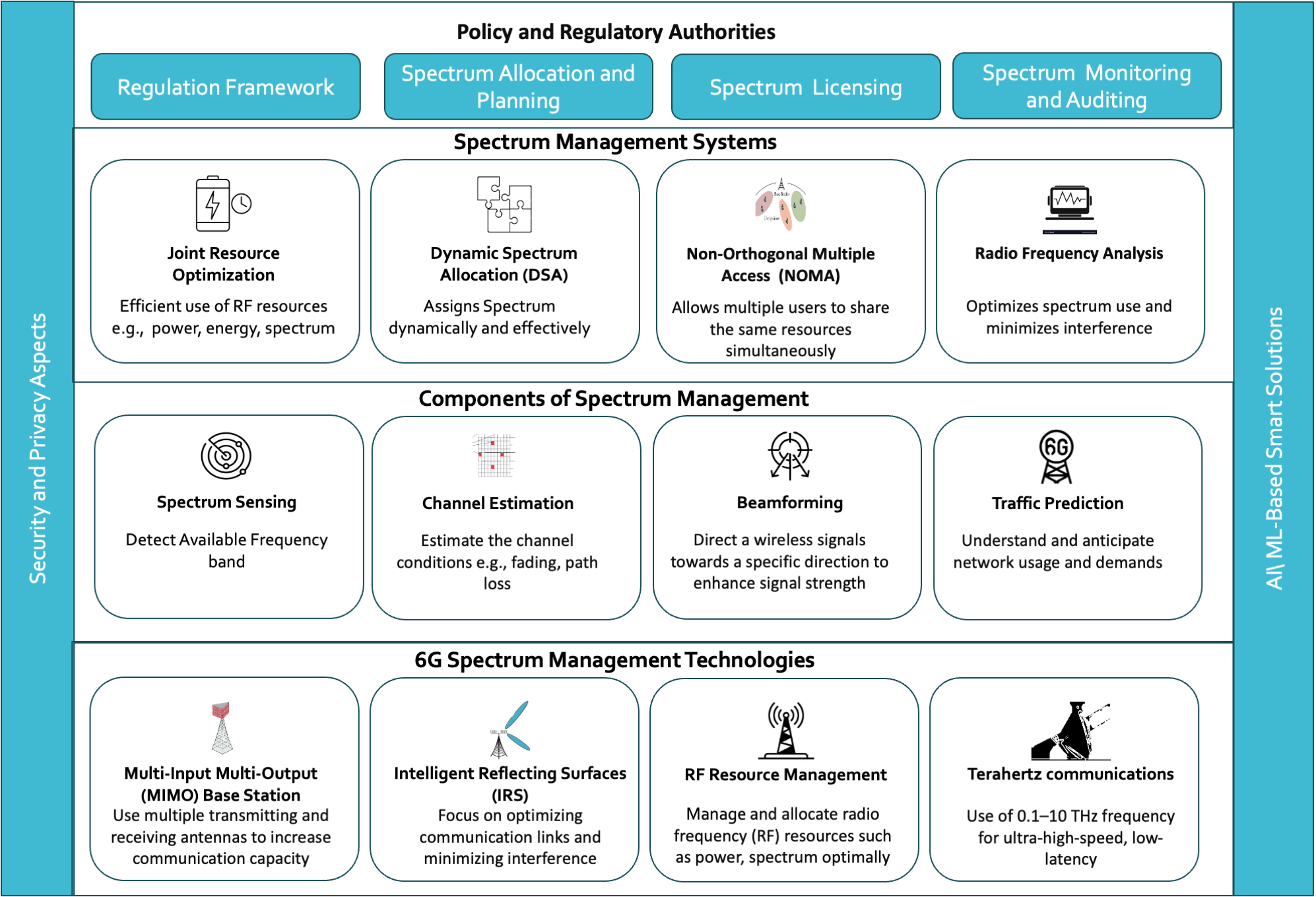}}
\caption{Spectrum Management Framework for B5G and 6G Networks}
\label{sm_framework}
\end{figure*}
 \subsection{Spectrum Management Framework for B5G and 6G Networks} 
 \label{componentsSM}
In this section, we provide a brief overview of key elements of the spectrum management process, acquainting the reader with the foundation of spectrum management. We discuss these elements using a layered approach within a spectrum management framework, as illustrated in Figure \ref{sm_framework}. This framework is organized into four distinct layers, each contributing to the comprehensive management of the spectrum.

\textbf{Layer 1: Policy and Regulatory Authorities}
At the foundation, this layer outlines the regulatory and policy framework governing spectrum management. It includes:
\emph{Regulation Framework:} Directed by international bodies such as the ITU, this framework facilitates a coordinated approach to spectrum management, ensuring efficient and equitable access \citep{itu2020spectrum2}.
\emph{Spectrum Allocation and Planning:} This process involves segmenting the spectrum into bands for various services and users, aligning with national and international agreements \citep{itu2020spectrum2, stine2004spectrum}.
\emph{Spectrum Licensing:} Regulatory bodies manage spectrum access through licensing, distinguishing between individual licenses, apparatus licenses, and unlicensed spectrum to ensure a balance between exclusive use and shared access \citep{itu2020spectrum}.
\emph{Spectrum Monitoring and Auditing:} This is crucial for policy enforcement, identifying unauthorized or inefficient use, and optimizing spectrum utilization.
\par\textbf{Layer 2: Spectrum Management Systems}
This layer describes the systems and frameworks for implementing spectrum management principles and technologies set forth by regulatory policies:
\emph{Dynamic Spectrum Allocation:} Systems for real-time spectrum reassignment based on current demands and conditions, aiming for spectral efficiency optimization \citep{itu2020spectrum}.
\emph{Joint Resource Optimization (JRO):} Frameworks for managing spectrum and other resources (e.g., power, infrastructure) to enhance network performance.
\emph{Non-Orthogonal Multiple Access (NOMA):} A technique for multiple users to share spectrum resources through power levels or spreading codes differences.
\emph{RF Analysis:} For monitoring, diagnosing, and planning spectrum usage to ensure regulation compliance and performance optimization.
\par\textbf{Layer 3: Components of Spectrum Management}
This layer delves into technical aspects and methodologies required by Spectrum Management Systems:
\emph{Spectrum Sensing and Cognitive Radio:} Technologies for real-time detection of spectrum use, enabling dynamic access and efficient sharing \citep{nist2022esc}.
\emph{Channel Estimation and Beamforming:} Methods to enhance signal quality and reduce interference by optimizing the propagation environment.
\emph{Traffic Prediction:} Using AI/ML to forecast network load and usage patterns, aiding in dynamic spectrum allocation and management.
\par\textbf{Layer 4: 6G Spectrum Management Technologies}
Lastly, emerging technologies are vital for enhancing spectrum efficiency in the era beyond 5G:
\emph{MIMO} is a B5G wireless technology that uses multiple transmitters and receivers to transfer more data simultaneously. \citep{ling2010measurement, merias2017study}.
\emph{Intelligent Reflecting Surfaces (IRS):} Complementing MIMO, IRS controls electromagnetic waves to optimize signal propagation and reception.
\emph{RF Resource Management:} This involves strategies for optimal spectrum use, including power control and user allocation, leveraging algorithms for efficient utilization \citep{oh2022general}.
\emph{TeraHertz Communication:} 6G spectrum management is an enabling technology that utilizes the untapped terahertz frequency band (0.3–10 THz) for ultra-high data-rate, low-latency wireless data transmission, addressing the bandwidth limitations of previous generations.

Across all layers, ensuring security and privacy in spectrum management processes and leveraging AI/ML for enhancing prediction, planning, real-time allocation, and enforcement are paramount.





 \section{Research Methodology} 
 \label{section-research-methodology}
 This section describes the protocol for the literature review, which includes establishing research questions, formulating a search strategy, specifying criteria for inclusion and exclusion, detaiiled the study selection, data extraction and synthesis process.

\subsection{Research Questions }
This literature review aims to systematically examine the existing research on AI-enabled spectrum management (AISM) systems, with a focus on their application beyond 5G (B5G) networks. The review will identify, analyze, and synthesize the current knowledge on AISM, addressing the learning paradigms employed, the datasets used for system training and evaluation, the performance metrics critical for assessing system effectiveness, and the security and privacy concerns specific to AISM in the context of 5G and B5G networks.
By exploring these dimensions, the review seeks to uncover gaps in the current literature, provide insights into the challenges and opportunities associated with AISM, and suggest directions for future research.
We achieve this by formulating and addressing two main research questions (RQs).

\textbf{RQ1: Exploration of AISM Systems in B5G Context
}
\begin{itemize}
    \item RQ1.1: What are the specific AISM systems developed or proposed for use in B5G networks, and how do these systems enhance spectrum management compared to traditional methods?
    \item RQ1.2: What are the primary learning paradigms (e.g., deep learning, reinforcement learning, etc.) employed in AISM for 5G and B5G, and how do they contribute to the systems' adaptability and efficiency?
    \item RQ1.3: What specific types of datasets (e.g., real-world spectrum usage data, synthetic datasets, etc.) are utilized for the training and evaluation of AISM systems in 5G and B5G networks, and what challenges are associated with these datasets?
    \item RQ1.4: What performance evaluation metrics (e.g., accuracy, efficiency, scalability, etc.) are critical for assessing the effectiveness of AISM systems in managing the spectrum for B5G?
\end{itemize}

\textbf{RQ2: Security and Privacy Concerns in AISM for Spectrum Management
}
\begin{itemize}
    \item RQ2.1: What new security and privacy threats are introduced by implementing AISM systems in B5G networks, and what existing threats in traditional spectrum management could potentially be adapted by AISM?
    \item RQ2.2: What strategies, techniques, or frameworks are being proposed or utilized to mitigate the identified security and privacy threats in AISM systems?
  
\end{itemize}

\subsection{Search Strategy}
\begin{figure*}
  \centering
  \includegraphics[width=.99\linewidth]{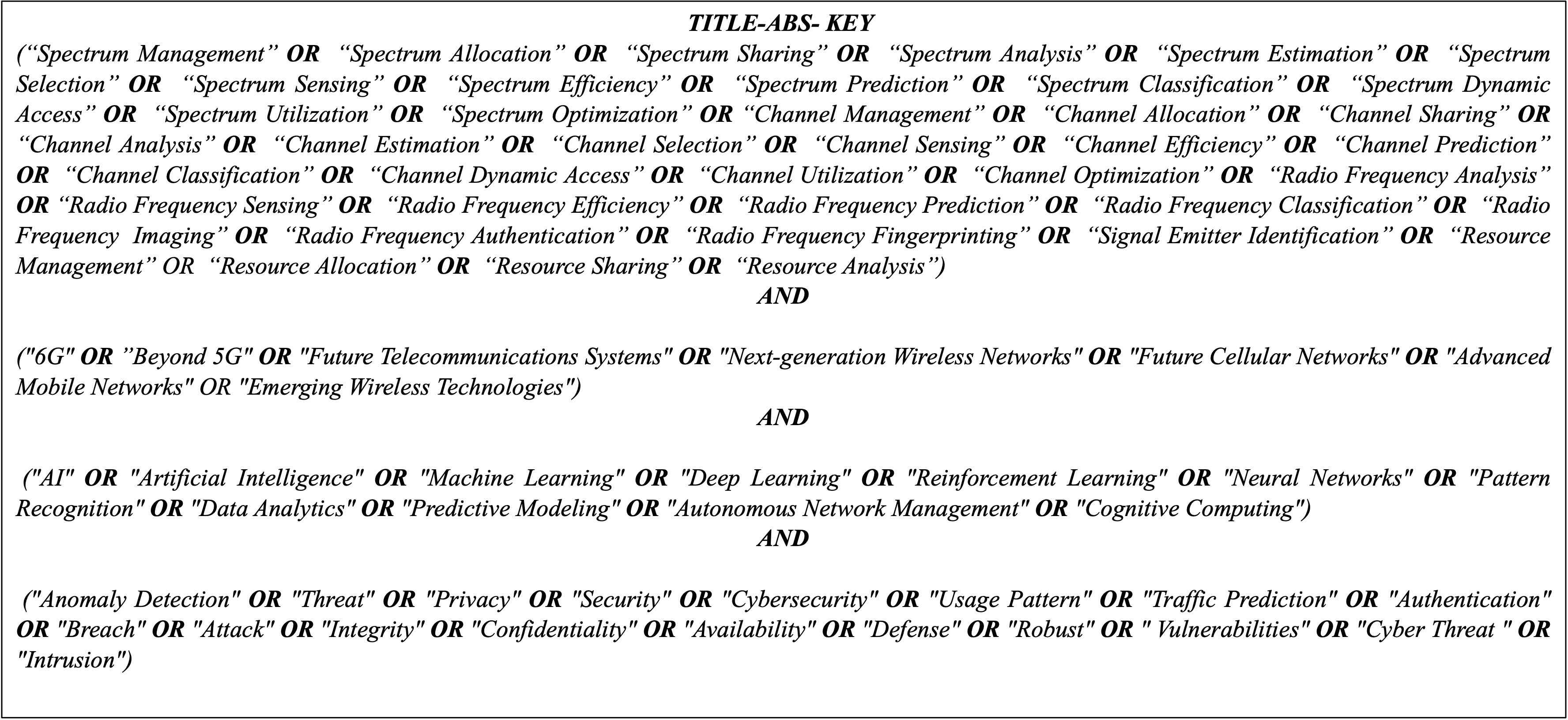}

\caption{Our Search String}
\label{Project4SearchString}
\end{figure*}

To systematically retrieve relevant papers, we followed the guidelines outlined in \citep{slrguidelines}. Our search strategy consisted of two main steps:
\begin{itemize}
    \item Database-driven Search: We initiated our search using a database-driven approach \citep{slrguidelines}. This method involved selecting an initial set of papers from reputable databases. Additionally, we supplemented this approach with forward and backward snowballing \citep{snowballingguidelines}.
    \item Search Term Formulation:
Radio Frequency characterization, 6G and B5G, AI, and Security. Logical operators such as AND and OR were employed to combine identified spectrum management components with relevant keywords related to AI, 6G, B5G, and security. Iterative refinement involved pilot searches on IEEE and ACM databases to align the string with our research focus.
Pilot searches revealed that certain keywords, like "resource optimization," yielded unrelated papers. To address this, a refined query was tested: ("Resource Optimization" AND ("MACHINE LEARNING" OR "DEEP LEARNING" OR "ARTIFICIAL INTELLIGENCE") AND (6G OR 5G) AND (SECURITY OR PRIVACY)). Upon analyzing the results, it was found that none of the returned papers met our inclusion criteria, leading to the exclusion of these terms from the final search string. Furthermore, quotation marks were strategically used to denote phrases, enhancing the precision of the search process. The complete search string is detailed in Figure \ref{Project4SearchString}.
\end{itemize}

\textbf{Data Sources}
We conducted searches in the five widely recognized digital databases \citep{computersciencetrendsDL}. The data sources utilized in this study are IEEE \footnote{\url{https://ieeexplore.ieee.org }}, 
 ACM  \footnote{\url{https://dl.acm.org/}} ,
Wiley \footnote{\url{ https://onlinelibrary.wiley.com }}, 
Science Direct \footnote{\url{ https://www.sciencedirect.com  }} and
Scopus \footnote{\url{ https://www.scopus.com }}.
   Our search string was specifically applied to the title, abstract, and keywords of the papers. To maintain consistency, we imposed a filter based on the publication year, exclusively considering papers published between 1st January 2013 and 1st September 2023.

\textbf{Inclusion and Exclusion Criteria} Table \ref{Project4InclusionCriteria} outlines the rigorous inclusion and exclusion criteria used to select papers for the study, ensuring a focus on high-quality, relevant research. The criteria were first tested on a preliminary group of ten papers, allowing for refinement based on initial findings. Key to the exclusion criteria was the requirement for papers to be published in venues with an H5-index above 60, according to Google Scholar, to guarantee a baseline of quality and influence in the field over the past five years. The study specifically targeted papers proposing AI-based approaches for spectrum management and security, thereby excluding those that focused on non-AI methodologies like blockchain, databases, or alternative technologies. This approach ensured the compilation of literature was both relevant to the study's AI focus and of a high academic standard.

\begin{table*}[bt!]
\footnotesize
\caption{Inclusion and Exclusion Criteria}
\label{Project4InclusionCriteria}
\centering
\resizebox{\textwidth}{!}{\begin{tabular}{|c|c|p{15cm}|}
\hline
    \textbf{Criteria} & \textbf{ID} & \textbf{Description}\\
    \hline
    \multirow{6}{*}{\textbf{Inclusion}} 
    & I1 & Studies using AI techniques (e.g., supervised, unsupervised, or anomaly-based detection).
    \\ \cline{2-3}

     & I2 & Studies that are peer-reviewed. \\ \cline{2-3}
     & I3 & Studies that report contextual data (i.e., ML model, feature engineering techniques, evaluation datasets, evaluation metrics and security angle). \\ \cline{2-3}
     & I4 & Papers from authors with high ($>60$) H5-Index. \\ \cline{2-3}
     & I5 &Papers from standardization bodies.\\ \cline{2-3}
     &I6& Primary studies will be included. \\
     \hline
    \multirow{8}{*}{\textbf{Exclusion}} & E1 & Studies that use other than AI techniques (e.g., access control mechanism, blockchain) for spectrum management. \\ \cline{2-3}
     & E2 & Studies that are not on physical or MAC layers (e.g., resource optimization in cloud computing). \\ \cline{2-3}
     & E3 & Short papers less than five pages. \\ \cline{2-3}
     & E4 & Non-peer reviewed studies, except for papers from leading organizations. \\ \cline{2-3}
     & E5 & Secondary studies, surveys, vision papers, and reviews.  \\ \cline{2-3}
     & E6 & Papers in languages other than English. \\ \cline{2-3}
     &E7& Papers are on Adhoc networks and do not apply to Cellular and Wifi networks. \\ \cline{2-3}
     &E8 & Studies that are published before 2013. \\ \hline
\end{tabular}}
\end{table*}
\begin{figure*}[tb!]
\centerline{\includegraphics[width=\textwidth]{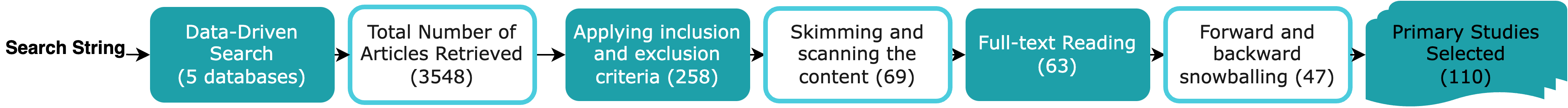}}
\caption{The SLR Process}
\label{slrprocess}
\end{figure*}

\begin{figure*}[tb!]
\centerline{\includegraphics[width=\textwidth]{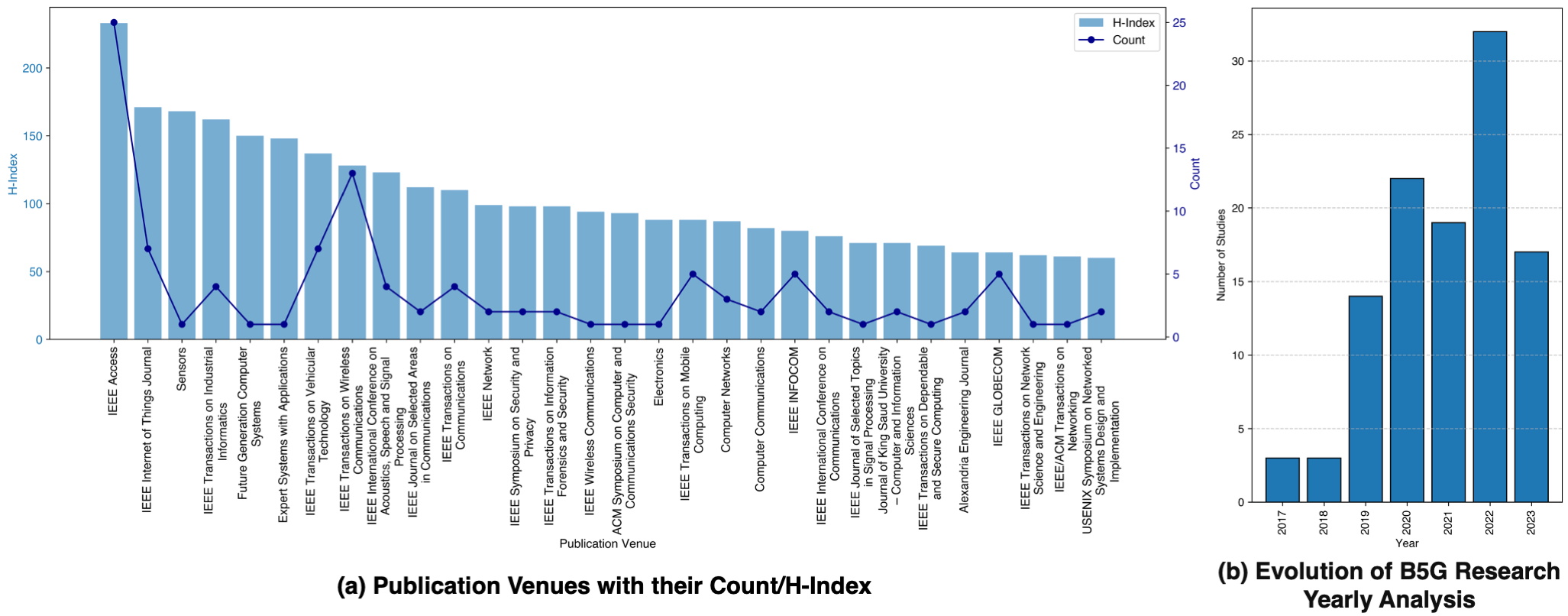}}
\caption{The Statistics of the Reviewed Studies}
\label{statics_studies}
\end{figure*}
\subsection{Selection of Primary Studies}
\label{3.5}
Fig~\ref{slrprocess} shows the steps we followed to select the primary studies. 
After following the complete process and repeatedly applying inclusion and exclusion criteria (Table~\ref{Project4InclusionCriteria}), 80 studies were selected for the review.
We initiated our research by examining each data source using a specific search string, identifying an initial set of 3548 papers. Notably, the Science Direct database contributed the highest number, totaling 1442 papers.
Upon obtaining the initial results, we implemented inclusion and exclusion criteria (see Table~\ref{Project4InclusionCriteria}, employing GPT-4 for title, abstract, and keyword-based filtering. Collaboratively, two authors cross-verified the selected papers to ensure alignment with the inclusion criteria, resulting in a refined set of 258 papers.
Further scrutiny involved a thorough review of the full text content, revealing that only 69 papers met our stringent inclusion and exclusion criteria. Subsequently, a detailed examination of these 69 papers identified a subset of 63 papers that directly addressed the focus of our study.
To enhance the comprehensiveness of our dataset, we employed both forward and backward snowballing techniques in accordance with guidelines outlined in \citep{RN64}. This iterative process led to the identification and inclusion of 47 additional papers. Consequently, a final selection of 110 papers was deemed suitable for the subsequent data extraction phase. Each of the primary studies analyzed in this work has been assigned an ID (S\#) that is used in the remainder of this SLR to distinguish them from other references. Please refer to the list of primary studies provided in Github repository \citep{Sabir2024SLR}.

Figure \ref{statics_studies} (a) shows the publication venue  and their H-Index of the selected studies. Notable venues include "IEEE Access" with the highest count of 25 papers and an impressive H-index of 233, indicating significant impact and influence. "Computer Networks" and "IEEE Transactions on Wireless Communications" also stand out with high H-index values of 87 and 128, respectively. Additionally, journals such as "Expert Systems with Applications" and "Future Generation Computer Systems" showcase noteworthy H-index scores of 148 and 150, highlighting their scholarly significance. The distribution of papers across various conferences and journals, each with distinct H-index metrics, provides a diverse perspective on the impact and reach of the selected publications in the field of engineering and computer science. Figure \ref{statics_studies} (b) provides the temporal trend of the studies since 2017.  In particular, there is a steady increase in the number of studies from 2019 to 2022, with a significant surge in 2022, where 26 studies were published. This spike suggests a recent upswing in research activity or a heightened interest in the subject matter during that year.
The data reveals a general upward trajectory, indicating a growth in the body of knowledge over time. The years 2022 and 2023 also exhibit considerable research output, with 32 and 17 studies, respectively.

\subsection{Data Extraction and Synthesis}
\label{3.6}
\subsubsection{Data Extraction}
The data extraction form was devised to collect data from the reviewed papers. Table~\ref{dataextractionform} shows the data extraction form. 
A pilot study with eight papers was done to judge the comprehensiveness and applicability of the data extraction form. The data extraction form was inspired by \citep{RN70} and consisted of two parts.
(i) Qualitative Data (D1 to D8): For each paper, a short critical summary was written to present the contribution and identify the strengths and weaknesses of each paper. The extracted demographic information included author information, h-index \citep{RN71}, conference or journal name, publisher name, year of publication and the title of the paper
Demographic information was extracted to ensure the quality of the reviewed papers. 
 (ii) Context Data (D11 to D22): Context of a paper included the details in terms of RQ1 and RQ2 were considered.
\begin{table*}[tb!]
\caption{Data type and item extracted from each study and related research questions enclosed in parenthesis}
\label{dataextractionform}
\centering
\footnotesize
\resizebox{\textwidth}{!}{\begin{tabular}{|c|c|c|l|}
    \hline
    \textbf{Data Type} & \textbf{Id} & \textbf{Data Item} &\textbf{Description} \\ \hline
    \multirow{10}{*}{Qualitative Data} & D1 & Title & The title of the study. \\
    \cline{2-4}
     & D2 & Author & The author(s) of the study. \\ \cline{2-4}
     & D3 & Venue & Name of the conference or journal where the paper is published. \\ \cline{2-4}
     & D4 & Year Published & Publication year of the paper. \\ \cline{2-4}
     & D5 & Publisher & The publisher of the paper. \\\cline{2-4}
     & D6 & Article Type & Publication type i.e., book chapter, conference, journal. \\ \cline{2-4}
     & D7 & H-index & Metric to quantify the impact of the publication. \\ \cline{2-4}
     & D8 & Summary & A brief summary of the paper along with the major strengths and weaknesses \\ \hline
    \multirow{6}{*}{Context (RQ1)} & D9 & Application & Application of Spectrum Management the study is considering. \\ \cline{2-4}
     & D10 & Dataset Type & Type of Dataset used by the study. \\ \cline{2-4}
     & D11 & Data Type & Type of Data used by the study. \\ \cline{2-4}
     & D12 & Tool or Testbed & What type of simulation tool or testbed to generate the dataset. \\ \cline{2-4}
     & D13 & Availability & Where the dataset is available publicly. \\ \cline{2-4}
     
     & D14 & AI Approach & Type of AI algorithm used. \\ \cline{2-4}
      & D15 & Model & Type of AI classifier or model used. \\ \cline{2-4}
     & D16 & Performance Metrics & The performance metrics study report its evaluation. \\ \hline
     
    \multirow{5}{*}{Context (RQ2)} & D17 & Type of Security and Privacy & Either paper covers security and privacy issues of SM or AISM  \\ \cline{2-4}
     & D18 & Type of Study & Does the study cover attack or defence? \\ \cline{2-4}
    & D19 & Threat type & What type of attack the paper focus on?\\ \cline{2-4}
     & D20 & Vulnerable type & What is the cause of the attack?\\ \cline{2-4}
      & D21 & Attacker Capability & What capability attacker needs to have to perform an attack?\\ \cline{2-4}
     & D22 &Defence & What type of defense the paper focus on? \\ \hline

\end{tabular}}
\end{table*}
\subsubsection{Data Synthesis}
 Each paper was analysed based on two types of data mentioned in Table~\ref{dataextractionform} as well as the research questions. 
The qualitative data (D1-D8) and context-based data (D11 to D22) were examined using thematic analysis \citep{RN72}.  
The thematic analysis was performed in five steps.
\begin{enumerate}
\item Data comprehensibility: In this step, data items (D8 to D22) were read and analyzed to form initial ideas regarding each RQ. 
\item Initial Codes generation: The initial patterns (e.g. type of data used by studies and ML classifier used) in the data were recognized, and initial codes, e.g., data type code (network, application, resource, and structured and unstructured), analysis type (statistical, structural, contextual) were assigned to each study. 
\item Searching for the theme: For each RQ, the previously extracted codes were grouped into clusters to form a potential theme.
\item Reviewing and refining themes: All the clusters were reviewed based on each RQ and refined to provide finer-grained and semantically coherent information regarding the primary studies. 
\item Defining and naming theme: lastly, clear and precise names for each theme were chosen and assigned.
\end{enumerate}

%
 \begin{table*}[!tb]
\centering
\resizebox{\textwidth}{!}{%
\begin{tabular}{|p{1cm}|p{1.5cm}|p{5cm}|p{6cm}|p{6cm}|p{3cm}|}
\hline
\textbf{AISM} & \textbf{Apps} & \textbf{Description} & \textbf{Strengths} & \textbf{Weakness} & \textbf{Studies} \\
\hline
\multirow{13}{*}{\rotatebox[origin=c]{90}{\parbox{3.5cm}{\textbf{Resource Mgt [38.18\%]}}}}  & \multirow{3}{*}{\rotatebox[origin=c]{90}{\parbox{2cm}{DSA [17.27\%]}}}
& A method where secondary users access unused frequency bands based on real-time spectrum availability, enhancing spectral efficiency. & - Maximizes the use of available spectrum.

\leavevmode\newline - Allows dynamic allocation of spectrum based on current needs. \leavevmode\newline -Reduces signal interference through intelligent spectrum allocation.  & - Complex to develop and integrate into existing systems.\leavevmode\newline - Inherits the security and privacy risk with AI decisions, data and models. \leavevmode\newline - Difficult to ensure compatibility with various network standards and devices. & [S2, S12, S15, S25, S27, S28, S37, S39, S40, S41, S43, S44, S45, S48, S52, S99, S100, S103, S104] \\
\cline{2-6}
 & 
 \multirow{4}{*}{\rotatebox[origin=c]{90}{\parbox{1.65cm}{Traffic Prediction [8.18\%]}}} & Estimating future network demand and usage to optimize resource allocation and spectrum utilization. & - Allocates resources proactively based on predicted demand.
\leavevmode\newline - Adapts to changing network conditions for optimal spectrum use.
\leavevmode\newline - Can reduce operational costs by anticipating network needs. & - Accurate predictions depend on quality data and algorithms.
\leavevmode\newline - Requires continuous tuning and maintenance.
\leavevmode\newline - Handling sensitive user data can raise privacy concerns.& [S46, S54, S57, S58, S59, S60, S61, S101] \\
\cline{2-6}
 & \multirow{4}{*}{\rotatebox[origin=c]{90}{\parbox{2cm}{JRO [13.63\%]}}} & Efficiently managing and distributing resources like spectrum, power, and bandwidth in high-performance networks. & - Improves overall network performance by optimizing multiple resources.
\leavevmode\newline - Utilizes historical data for better resource distribution.
\leavevmode\newline - Dynamically adjust to changing network environments and demands. & - Implementing JRO systems can be complex and expensive.
\leavevmode\newline - Must ensure secure and reliable operations.
\leavevmode\newline - Effectiveness relies heavily on data availability and quality. & [S7, S8, S9, S10, S11, S16, S21, S22, S23, S26, S34, S50, S51, S55, S106] \\
 \hline
\multirow{16}{*}{\rotatebox{90}{\textbf{Beam Mgt [29.09\%]}}} & \multirow{6}{*}{\rotatebox[origin=c]{90}{\parbox{2cm}{Beam Selection [18.18\%] }}}& Directing signal transmission in specific directions in high-frequency bands to optimize spectrum usage. & - Ensures highly efficient connections through detailed data analysis.
\leavevmode\newline - Quickly adjusts beam direction in response to environmental changes and user mobility.
\leavevmode\newline - Enhances network throughput and reduces signal interference by selecting suitable beams.
 & - Demands sophisticated significant resources, increasing infrastructure complexity.
\leavevmode\newline - Relies heavily on the availability and quality of data.
\leavevmode\newline - Raises privacy and security concerns.
\leavevmode\newline - Difficulties in integrating with existing systems, requiring extensive compatibility efforts. & [S62, S63, S65, S66, S67, S68, S69, S71, S74, S75, S76, S79, S81, S82, S83, S84, S91, S92, S96, S98] \\
\cline{2-6}
 &\multirow{5}{*}{ \rotatebox[origin=c]{90}{\parbox{1.75cm}{Beam Alignment [2.72\%]}}} & Adjusting directional antennas for optimal transmitter-receiver alignment, crucial in beamforming and MIMO. & - Ensures optimal transmitter-receiver connections.
\leavevmode\newline -  Improves overall signal quality and network performance.
\leavevmode\newline -  Minimizes signal interference with other users.
&
- Requires high precision, which can be difficult to maintain.
\leavevmode\newline - Demands significant processing power.
\leavevmode\newline - Risks of alignment errors affecting network performance. & [S64, S89, S102] \\
\cline{2-6}
 & \multirow{4}{*}{\rotatebox[origin=c]{90}{\parbox{1.60cm}{Beam Optimization [5.45\%]}}} & Dynamically adjusting directional beam parameters to maximize signal strength and communication efficiency. & - Maximizes signal strength for better communication.
\leavevmode\newline - Adjusts to physical and network changes.
\leavevmode\newline - Supports more users and data throughput. & - Requires sophisticated algorithms.
\leavevmode\newline - Effectiveness depends on comprehensive data input.
\leavevmode\newline - Difficult to integrate with existing network & [S19, S47, S70, S77, S80, S90] \\
\cline{2-6}
 & \multirow{4}{*}{\rotatebox[origin=c]{90}{\parbox{1.5cm}{Beam Tracking [2.72\%]}}} & Maintaining directional beam links between transmitters and receivers in high-frequency communication systems. & - Maintains stable connections in mobile environments.
 \leavevmode\newline - 
 Improves overall network reliability and speed.
 \leavevmode\newline - 
Keeps track of user movements for optimal service.& - Demands high computational power.
\leavevmode\newline - Must process data quickly to maintain tracking.
\leavevmode\newline - Physical obstructions can disrupt tracking accuracy & [S72, S73, S78] \\
\hline

\multirow{8}{*}{\rotatebox[origin=c]{90}{\parbox{2.75cm}{\textbf{Channel Mgt [23.63\%]}}}} & \multirow{5}{*}{\rotatebox[origin=c]{90}{\parbox{1.5cm}{Channel Estimation [13.63\%]}}} & Accurately determining channel characteristics in wireless systems, vital for decoding signals in advanced networks. & - Identifies and utilizes available spectrum bands effectively.
\leavevmode\newline - Minimizes interference with existing spectrum users.
\leavevmode\newline - Facilitates harmonious spectrum sharing.& - Complexity in High-Frequency Bands\leavevmode\newline - Data Requirement for Accuracy\leavevmode\newline - Processing Delays & [S3, S5, S13, S14, S18, S35, S53, S56, S93, S95, S105, S107, S108, S109, S110] \\ 
\cline{2-6}
 & \rotatebox[origin=c]{90}{\parbox{1.5cm}{Spectrum Sensing [10\%]}} & Identifying used and unused frequency bands in wireless communications for efficient spectrum utilization. & - Efficient Spectrum Usage \leavevmode\newline - Reduced Interference\leavevmode\newline - Better Coexistence with Existing Users & - Sensing Accuracy Challenges \leavevmode\newline - High Computational Demand\leavevmode\newline - Dynamic Environment Adaptation Issues & [S24, S29, S30, S31, S32, S36, S38, S42, S49, S94, S97] \\ 
 
 \hline
\end{tabular}}
\leavevmode\newline
\caption{Classification of AI-enabled Applications in Spectrum Management (AISM): Strengths, Limitations, and Studies [Management (Mgt), Percentage of reviewed studies is given in square brackets]}
\label{table:spectrum_management_ai}
\end{table*}

\section{RQ1: AI-Enabled Spectrum Management (AISM)}
 \label{section_RQ1}
In this section, we present a multifaceted review and analysis of AI-enabled spectrum management applications. Our analysis, centered on data items D9–D16 as detailed in Table ~\ref{dataextractionform} to answer this RQ. This question sheds light on four main dimensions: firstly, we analyze, classify, and map the reviewed studies by their spectrum management approaches; secondly, we delve deeper into the learning methodologies utilized in these studies; thirdly, we examine the datasets employed by the reviewed studies; and finally, we provide a synopsis of the performance metrics used in these methodologies.

\subsection{RQ1.1: Classification of AISM Applications }
Table \ref{table:spectrum_management_ai} presents a classification of AI-enabled Spectrum Management applications, their description, strengths, limitation and provides a mapping to the reviewed studies (\citep{Sabir2024SLR}).
ASIM approaches fall into three main categories: Resource, Beam, and Channel Management.

\subsubsection{Resource Management} 
Resource management in B5G involves the dynamic allocation of available network resources at physical layer, such as energy, spectrum and computational resources. 
In B5G networks, resource management increasingly adopts AI methods, as evidenced by 39.78\% of the studies analyzed. This shift is primarily due to the limitations of traditional approaches, which rely on static rules and predefined protocols and are inadequate for the dynamic nature of B5G environments. In contrast, AI offers real-time adaptability to evolving network conditions and user demands. We further classified the studies under three sub-areas: Dynamic Spectrum Allocation (DSA), Traffic Prediction, and Joint Resource Optimization (JRO).
AI-driven techniques are increasingly adopted in \emph{DSA:} due to the limitations of traditional static frequency management in modern networks, as indicated by 16\% of the studies reviewed. These AI methods aim to enhance efficiency in areas like 6G and IoT, focusing on faster network convergence, reduced energy use, and improved security. For instance, study [S2] proposed an AI-based dynamic bandwidth allocation method that adapts to the current needs of each cell, optimizing frequency resource usage with a Utilization Factor (UF) close to 1 for high efficiency. This intelligent approach not only serves more traffic, especially during peak hours but also alleviates congestion in 9\% - 12\% of cells by using the same level of system bandwidth more effectively compared to traditional static methods.
Within these DSA studies, a few of them also addressed the security and privacy concerns in AI-enabled DSA. For example, [S44] introduces a DSA approach that use robust security protocols to ensure incumbent protection without centralized coordination. The paper introduces two-step AI-based algorithm that can recognize, learn, and predict the incumbent's transmission patterns with 95\% accuracy in near real-time.
\emph{Traffic Prediction:}, comprising 9\% of the studies in AI-enabled resource management, serves multifaceted purposes. Notably, AI is leveraged for precise traffic forecasting. For instance, studies [S57, S58] employ federated learning and data augmentation to effectively anticipate future network demands, showcasing improvements in spectrum optimization and allocation.
Beyond traffic forecasting, AI plays a crucial role in fortifying cybersecurity for spectrum management. For example, [S46] introduces a Sequential Deep Reinforcement Learning Algorithm (SDRLA) to identify jamming patterns and make real-time channel selections. The approach demonstrated effectiveness by achieving up to 50\%-70\% reduction in channel switching frequency and maintaining throughput between 4-7 in dynamic jamming environments, showcasing its adaptive response to changing jamming patterns without prior information. 
\emph{JRO:} Notably, our SLR highlighted the prevalence of this theme, with 15\% of studies aligning with it. These studies mainly use AI-driven JRO to provide network resource to end-users by minimizing the overall latency and energy consumption. For example, [S7] proposed a joint resource management and computation offloading scheme by integrating deep reinforcement learning in Cyber-twin-enabled 6G wireless networks.  
Another interesting study in [S21] where the authors suggest a Device to Device (D2D)-aided federated learning scheme with matching theory-based incentives and channel forecasting for collaborative learning optimisation.


\subsubsection{Beam Management}
Beam Management refers to managing the direction and shape of antenna beams, particularly in Massive MIMO (Multiple Input Multiple Output) systems in 5G and B5G wireless networks.
In B5G networks, 29.09\% of our studies highlight a transition to AI in beam management, moving away from traditional methods reliant on fixed algorithms and parameters. This shift is crucial in mmWave and massive MIMO environments, where AI-enabled beam management employs machine learning for adaptive, real-time decision-making. This approach significantly improves beam selection, alignment, and optimization, enhancing signal quality and reducing interference, particularly in mobile scenarios like vehicular networks. Consequently, AI-driven beam management elevates network performance and efficiency. We've categorized AI applications in beam management into four sub-areas: Beam Selection, Alignment, Optimization, and Tracking, with further details and study correlations presented in Table~\ref{table:spectrum_management_ai}. \emph{Beam Selection:}, accounting for 18.18\% of studies in our review, predominantly features AI applications in mmWave vehicular communications and massive MIMO systems. Key insights include:
mmWave Beam Selection: AI is leveraged for efficient beam selection in mmWave links, with approaches like a double-layer online learning algorithm and machine learning schemes [S62, S63, S65] to reduce overhead.
Moreover, AI aids in enhancing data rates and reliability in massive MIMO mmWave systems. Notable examples are Zhao et al.'s [S71] two-step neural network for beamforming, Kim et al.'s [S69] deep learning-based resource allocation, and [S81]'s trainable projected gradient detector for overloaded MIMO channels.

\emph{Beam Alignment:} Interestingly, we found only three studies in our reviewed literature that discussed Beam Alignment using AI.  In [S64], a beam alignment framework is introduced that leveraged visual information, integrating 3D object detection and a deep neural network for optimal beam pair inference in transceivers. Another study  [S89] presented a blind beam alignment method using deep reinforcement learning. This method achieved higher data rates without introducing additional overhead. The neural network employed handled both continuous and discrete actions, utilising RF fingerprints from base stations. 
\emph{Beam Tracking:} Only three investigations leveraging AI for beam tracking, two of them focused on mmWave communication. For example, [S72] presented online reinforcement learning for beam tracking and rate adaptation in millimetre-wave systems. The proposed beam tracking approach exhibited high throughput gains as compared to conventional strategies by 25-35\%. Similarly, [S73] presents a deep learning coordinated beam alignment solution that overcomes the challenges related to coverage, reliability, and hand-off latency by integrating machine learning to predict beamforming vectors at base stations. 
\emph{Beam Optimization:} The second most explored theme in beam management, is highlighted by six key studies. A notable contribution comes from [S19], which introduces a learning framework for RIS-assisted up-link user-centric cell-free systems. Leveraging a digital twin, this framework optimizes access point and user association, power control, and RIS beamforming. In a different approach, [S90] addresses the challenge of closing the efficiency gap between analogue and digital beamforming at mmwave frequencies. Their exploration of refined beam selection using compressed sensing and machine learning demonstrates considerable optimization over traditional methods. 

\subsubsection{Channel Management}
Channel Management in wireless networks involves handling the characteristics and dynamics of communication channels to optimize the transmission and reception of signals. 
Traditional channel management struggles with complexities in communication systems due to its reliance on linear models. In contrast, AI-enabled approaches, utilized in 23.67\% of reviewed studies, offer adaptive, data-driven solutions. These enhance accuracy and efficiency, particularly in handling non-linearities, low signal-to-noise ratios, and real-time demands. We sub-classify the channel management into two paradigms: Channel Estimation and Spectrum Sensing.
\emph{Channel Estimation:} Encompasses 13.63\% of the reviewed studies. Among them [S5,S33, S13, S14, S93] leveraged deep learning for accurate channel estimation. Whereas, most of the studies (12) 
were focused on the security of channel estimation against physical and adversarial attacks. For instance, Chen et al. [S79] explored the channel estimation problem from a jamming perspective. They presented a game-theoretic model using deep reinforcement learning for anti-jamming in Mobile Edge Computing networks. On the other hand, [S56] addresses adversarial threats to channel estimation models in Next-Generation wireless networks, proposing a defensive distillation-based mitigation method to enhance security. 
\emph{Spectrum Sensing:}
Our review highlights 10\% of studies focusing on spectrum sensing, predominantly employing deep learning for enhanced detection and adaptability. Key contributions include [S29]'s use of deep convolutional neural networks (CNNs) with transfer learning for efficient TV signal detection across varied environments, demonstrating significant improvements in accuracy and training efficiency. Similarly, [S38] adopts deep learning via CNNs to refine noise estimation for superior RF signal detection, illustrating deep learning's advantage over traditional methods in spectrum sensing.

\begin{figure}[!tb]

\centerline{\includegraphics[width=\textwidth]{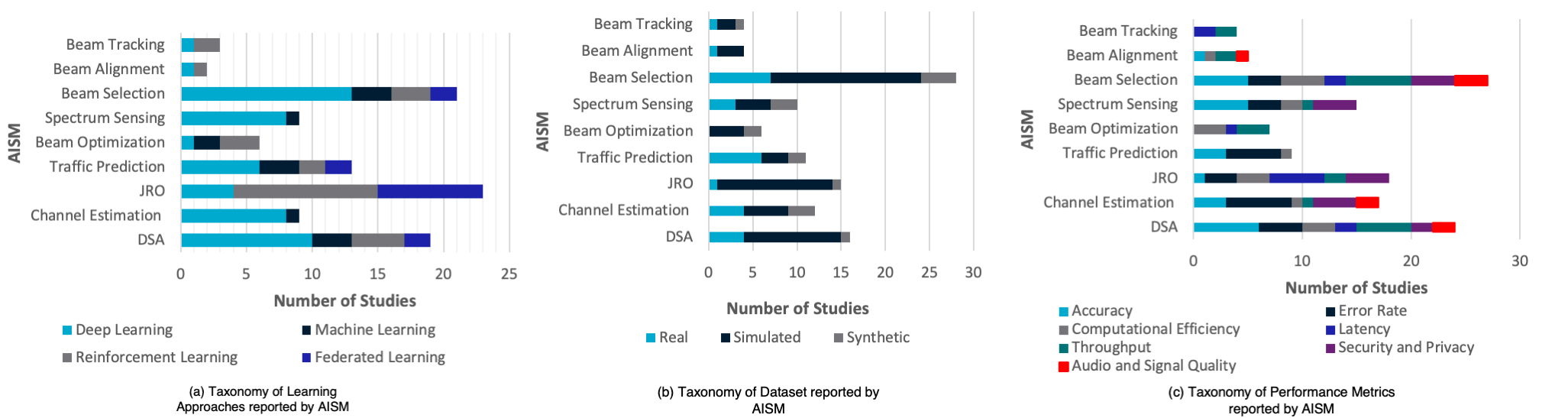}}
 \caption{Statistical distribution of reviewed studies over the devised taxonomies}
\label{Overall_Stats_RQ1}
 \end{figure}

\begin{table}[tb!]
\centering
\caption{Taxonomy of Learning Approaches in AISM for B5G Networks}
\label{tab:learning_approaches}
\footnotesize
\resizebox{\textwidth}{!}{
\begin{tabular}{|p{0.75cm}|p{2.5cm}|p{6cm}|p{6cm}|}
\hline
\textbf{Class} & \textbf{Sub-Class} & \textbf{Strengths} & \textbf{Weaknesses} \\
\hline
 \multirow{10}{*}{\rotatebox[origin=c]{90}{\parbox{1.65cm}{\textbf{Deep Learning}}}} & \multirow{6}{*}{\parbox{1.65cm}{N/A}} & - Can handle the dynamic, complex B5G spectrum environment, enhancing tasks like beam alignment and spectrum sensing. \leavevmode\newline - Optimizes spectrum management in data-intensive B5G networks. \leavevmode\newline - Superior in tasks such as beam selection, outperform traditional methods with higher signal detection accuracy. \leavevmode\newline - Offer automation, significantly reducing overhead in spectrum management. & - Opaque nature of these models complicates their interpretability and trustworthiness. \leavevmode\newline - Computational and energy requirements pose deployment challenges. \leavevmode\newline - High-quality training data is required, which is difficult to obtain for B5G scenarios. \leavevmode\newline - Vulnerable to adversarial attacks; ensuring DL model robustness is essential for reliable spectrum management. \\
\hline
\multirow{14}{*}{\rotatebox[origin=c]{90}{\parbox{1.65cm}{\textbf{Machine Learning}}}} & \multirow{3}{*}{\parbox{1.65cm}{Classification-based}} & - High accuracy and efficiency. \leavevmode\newline -  Adaptability to different network conditions. \leavevmode\newline -  Robust against attacks and interferences. & \leavevmode\newline - Requires significant labeled data. \leavevmode\newline - Risk of model complexity and overfitting. \leavevmode\newline -  Limited performance in complex and dynamic B5G scenarios. \\
\cline{2-4}
& \multirow{3}{*}{\parbox{1.65cm}{Clustering}} & - Efficient in unsupervised learning scenarios. \leavevmode\newline -  Effective for non-uniform channel distributions. \leavevmode\newline - . Useful in large data scenarios like Channel State Information (CSI) in MIMO. & - Sensitive to initial centroid selection. \leavevmode\newline -  Difficulty in choosing the number of clusters. \leavevmode\newline -  Limited to numeric data. \\
\cline{2-4}
& \multirow{3}{*}{\parbox{1.65cm}{Probabilistic Modelling}} & - Capable of handling uncertainty.\leavevmode\newline -  Effective in optimizing complex processes like beamforming. \leavevmode\newline -  Offers near-optimal performance with reduced complexity. & -  Development complexity. \leavevmode\newline -  Performance dependent on data distribution assumptions. \leavevmode\newline -  Potential scalability issues. \\
\cline{2-4}
& \multirow{3}{*}{\parbox{1.65cm}{Anomaly Detection}} & -  Improves network security by identifying unusual patterns. \leavevmode\newline - Ensures data integrity and reliability. \leavevmode\newline - Adapts dynamically to new anomalies. & - Risk of high false alarm rate. \leavevmode\newline - Dependence on effective feature selection. \leavevmode\newline - Limited by historical data for new anomaly types. \\
\hline
\multirow{10}{*}{\rotatebox[origin=c]{90}{\parbox{1.65cm}{\textbf{Federated Learning}}}} & \multirow{8}{*}{\parbox{1.65cm}{N/A}} & - Enhances data privacy and security by processing data locally. \leavevmode\newline -  Optimizes resource allocation and energy consumption, crucial for B5G networks. \leavevmode\newline - Scalable and flexible, handling a large number of devices and applications. \leavevmode\newline -  Adaptable to dynamic conditions and diverse user demands in 6G networks. \leavevmode\newline - Reduces latency and network load by processing data locally. & -  Model synchronization across diverse devices can be challenging. \leavevmode\newline - Communication overhead for model aggregation, significant in large-scale networks. \leavevmode\newline - Device heterogeneity impacts model training and convergence in diverse 6G environments. \leavevmode\newline - Complexity in implementing advanced learning techniques within FL. \\
\hline
\multirow{10}{*}{\rotatebox[origin=c]{90}{\parbox{1.65cm}{\textbf{Reinforcement Learning}}}} & Deep Reinforcement Learning & - Highly effective in dynamic spectrum allocation and resource management. \leavevmode\newline -  Can optimize network parameters for enhanced Quality of Service (QoS). \leavevmode\newline -  Adaptable to varying network conditions and user demands. & \leavevmode\newline - High computational demand can limit real-time applications. \leavevmode\newline -  Requires extensive training data, which may not always be available. \leavevmode\newline -  Complexity in model design and tuning can delay deployment. \\
\cline{2-4}
& Optimisation-based RL & -  Efficient in managing spectrum under rapidly changing conditions. \leavevmode\newline -  Suitable for scenarios with limited computational resources. \leavevmode\newline -  Can improve long-term spectrum utilization efficiency. & - May not perform as well in highly complex network scenarios. \leavevmode\newline -  Requires careful tuning of exploration-exploitation balance. \leavevmode\newline -  Potentially less effective than DRL in multi-objective optimization tasks. \\
\hline
\end{tabular}
}
\end{table}
 
\begin{table}[htbp]
\centering
\small
\caption{Learning Algorithms and Classification Approaches in AI-enabled Spectrum Management}
\label{tab:spectrum_management_LA}
\footnotesize
\begin{tabular}{|>{\raggedright\arraybackslash}m{6mm}|m{15mm}|m{15mm}|m{85mm}|}
\hline
\textbf{AISM} & \textbf{Sub-Type} & \textbf{AI Taxonomy} & \textbf{Classifier} \\
\hline
\multirow{14}{*}{\rotatebox[origin=c]{90}{\parbox{2cm}{\textbf{Resource Management}}}}
 & \multirow{5}{*}{{\parbox{1.65cm}{{DSA}}}} & DL (7) & CNN [S15,S39,S43,S44], CNN-RNN [S37,S100], MLP [S12,S28], Spatial Transformer Networks [S40], Multiple [S15], Auto-regressive Neural Network (NARNET) [S2], Residual networks (RNs) [S39]\\ \cline {3-4}
&&
DRL (3) & DDNN [S48], DQN [S41,S28]\\ \cline {3-4}
&&DL-FL(2) & RNN and Echo State Network (ESN) [S52,S25, S100]\\ \cline {3-4}
&&CB-ML (1) & SVM [S45], RF [S100], Adaboost [S100]\\ \cline {3-4}
&&Clustering (1)& K-means [S25]\\ \cline {3-4}&&PM (1) & Fuzzy Inference [S27] \\
\cline{2-4}
& \multirow{4}{*}{{\parbox{1.5cm}{{Traffic Prediction}}}} & DL (3) &  CNN [S59], LSTM and GAN [S57]\\ \cline {3-4}
&&DRL (1) & CNN [S46]\\ \cline {3-4}
&&DL-FL (1) & Multiple [S58]\\ \cline {3-4}
&& PM (1) & GMM [S61]\\ \cline {3-4}
&& AD (2) & Unsupervised AD [S54], Bayesian [S59] \\
\cline{2-4}
& \multirow{4}{*}{{\parbox{1.65cm}{{JRO}}}}& DL (2) & MLP [S34, S107]\\ \cline {3-4}
&&DRL (6) & DDPG [S106], Actor-Critic Policy Gradient [S50, S106], DQN (MLP) [S8, S11] Delayed DDPG [S7]\\ \cline {3-4}
&&DL-FL (8) & RNN and ESN [S21], CNN [S26]\\ \cline {3-4}
&&RL-FL(4) & Twin-Delayed DDPG [S16], Asynchronous Advantage Actor-Critic [S51, S106], DDPG [S23], DQN [S9, S10] \\
\hline

\multirow{4}{*}{\rotatebox[origin=c]{90}{\parbox{1.8cm}{\textbf{Channel Management}}}}
 & \multirow{2}{*}{{\parbox{1.65cm}{{Spectrum Sensing}}}}& DL (9) & Auto-encoder [S49], CNN [S24,S29,S36,S38,S42,S94], MLP [S32], CNN-RNN [S30]
\\ \cline {3-4}
&&CB-ML (1) & Naive Bayes (NB), SVM [S31] \\
\cline{2-4}
& \multirow{2}{*}{{\parbox{1.65cm}{{Channel Estimation}}}} & DL (9) & CNN [S5,S56,S93], LSTM [S56], MLP [S14], Encoder-decoder based RNN [S3, S18], meta-learning-based DL [S13], GAN [S95, S109]\\ \cline {3-4}
&&CB-ML (1) & SVM, DT, KNN, Boosted Trees [S35] \\
\hline
 \multirow{15}{*}{\rotatebox[origin=c]{90}{\parbox{1.75cm}{\textbf{Beam Management}}}}

& \multirow{2}{*}{{\parbox{1.65cm}{{Beam Alignment}}}}& DL (2) & CNN and LSTM [S64], Auto-encoder [S102]\\ \cline {3-4}
&&DRL&Deep Deterministic Policy Gradient (DDPG) [S89] \\
\cline{2-4}
&  \multirow{5}{*}{{\parbox{1.65cm}{{Beam Optimisation}}}} 
& DL (1) & CNN [S70]\\
\cline {3-4}
&&Clustering (1)&K-mean [S80]\\ \cline{3-4}
&&DRL (3) &Twin-delayed (DDPG) [S19], Deep Dueling Neural Network (DDNN) [S47], Proximal Policy optimisation (PPO) [S77]\\
\cline{3-4}
&&PM (1) & Dictionary learning Iterative Soft Thresholding (DLIST) [S90] \\
\cline{2-4}
&  \multirow{4}{*}{{\parbox{1.65cm}{{Beam Selection}}}} & DL (11) & CNN [S68,S76], LSTM [S74], CNN-LSTM [S65], MLP [S66,S67,S71,S81,S82,S83], Transfer Learning (TL) [S67]\\ \cline {3-4}
&&FL (2)& CNN [S96,S98]\\ \cline {3-4}
&&DRL (2) & Q-Learning [S69, S75]\\ \cline {3-4}
&&PM (1) & Cross-Entropy optimisation (CEO) [S79]\\ \cline {3-4}
&&ORL (1)&Multi-Armed Bandit (MAB) [S62]\\ \cline {3-4}
&&CB-ML (2) & SVM [S63, S91], KNN [S91] \\\cline {3-4}
\cline{2-4}
&  \multirow{4}{*}{{\parbox{1.65cm}{{Beam Tracking}}}}& DL (1) & CNN [S73]\\
&&DRL (1) & DDNN [S78]\\\cline {3-4}
&&ORL (1) & Adaptive Thompson sampling (ATS) [S72] \\
\hline

\end{tabular}
\end{table}

\subsection{ RQ1.2: Taxonomy of Learning Approaches}
In this section, we provide a comprehensive taxonomy of learning approaches employed in studies related to AISM. The taxonomy of learning approaches within the realm of AI-enabled spectrum management, along with their strengths and weaknesses, is meticulously presented in Table ~\ref{tab:learning_approaches}, whereas Table~\ref{tab:spectrum_management_LA} provides a mapping of learning approach (sub-type) with the classifiers reported and reviewed studies. Moreover, Figure \ref{Overall_Stats_RQ1} (a) illustrates the distribution of the learning approaches across various AISM solutions.

\subsubsection{Deep Learning (DL):} Use of DL is reported in 47.27\% of the reviewed studies, particularly for beam selection, Dynamic Spectrum Access (DSA) and Traffic Prediction as depicted in Figure \ref{Overall_Stats_RQ1} (b). These studies use notable neural network architectures like Long Short-Term Memory (LSTM) and CNNs. For example, [S74] showcases an LSTM for beam selection in massive MIMO mmWave communications. 
Whereas, DL's application in JRO is minimal, highlighted by a two studies [S34, S107], possibly due to JRO's complex and multi-stage nature accounting for multiple variables and constraints simultaneously.
Interestingly, Generative adversarial networks (GANs) has been used in two channel estimation studies [S94, S109]. For instance, study [S109] leverages GANs and a Siamese network in novel machine learning frameworks to detect spoofing attacks in 5G mmWave networks with high accuracy, using the unique SNR traces from IEEE 802.11ad devices. These methods show promising security enhancements for next-gen wireless networks, achieving up to 99\% detection accuracy.
Among different DL architectures, CNN and MLP emerge as the most prominent, reported in 19.35\% and 11.95\% of the reviewed studies, respectively.
\subsubsection{Machine Learning (ML)}
 Machine Learning is not commonly employed in AISM applications, with only 11.82\% of studies reporting its usage. It is primarily utilised in traffic prediction,  beam selection and DSA approaches
\emph{Classification-based ML (CB-ML)}
CB-ML operate on supervised learning methods, necessitating labeled data for training classifiers such as Support Vector Machine (SVM) and K-Nearest Neighbours (KNN). It is only adapted by 4.46\% reviewed studies. For example, [S63] employs an SVM classifier for efficient and rapid analogue beam selection in mmwave vehicular ad hoc networks (mmWave VANETs) with the goal of enhancing spatial spectrum reuse. The proposed ML approach utilises an iteration sequential minimal optimisation training algorithm to train data samples for all V2V links. The results demonstrate a higher performance than traditional solutions based on explicitly estimated channels.
\emph{Clustering} is also deployed in two study [S80, S25], for beam optimisation and DSA respectively. In [S80], a novel K-means clustering-based codebook design for MIMO communication systems was developed, leveraging large CSI datasets to form clusters whose centroids define the codebook. This method outperforms traditional designs by better accommodating non-uniform channel distributions, as confirmed by simulations. Whereas [S25] 
 utilizes K-means clustering for proactive caching in cellular networks, addressing privacy concerns and indirectly supporting spectrum management by optimizing resource use and ensuring data privacy.
\emph{Probabilistic Modelling (PM)} is utilized in three studies for handling data uncertainty: [S61] uses Gaussian Mixture Models (GMMs) for sensor data analysis, [S79] applies cross-entropy optimization (CEO) for efficient hybrid beamforming in MIMO systems, achieving 98\% efficacy of exhaustive searches with less complexity. \emph{Anomaly Detection (AD)} is explored for traffic prediction and data integrity in smart cities and IIoT for 6G networks. [S60] introduces a Bayesian system for accurate anomaly detection in sensor networks, and [S54] employs a novel algorithm with a multidimensional relationship diagram for improved anomaly detection in IIoT, enhancing reliability and interference handling.
\subsubsection{Federated Learning (FL)}
Federated Learning (FL) is a decentralised machine learning approach widely embraced in the AISM community (12.73\%), notably in JRO, DSA, and Traffic Prediction. For example, the [S21] study proposes an FL approach to enhance 6G networks, focusing on AI integration for JRO and addressing device privacy and channel variability. It introduces an incentive mechanism for user participation and employs echo-state networks for channel prediction, optimizing resource use and improving efficiency by reducing the overall time spent to complete the model training by 10\%, energy consumption by 5\%, increase the accuracy and revenue (as a function of the number of devices) by 10\% and 15\% respectively in 6G environments.
Another variant of FL is 
\emph{Deep-learning based Federated Learning (Deep FL)}. It
 applies deep neural networks to decentralized training while ensuring data privacy. 
[S26] addresses the challenge of high power consumption in Federated Edge Learning (FEEL) by utilizing CNNs with Non-Orthogonal Multiple Access (NOMA) in 5G networks, significantly outperforming existing models. This approach effectively reduces energy usage during the processing and transmission of data, enhancing the efficiency and sustainability of these systems.
 Another research [S58] applies FL to 5G cellular traffic prediction, employing various neural networks like RNN, CNN, LSTM, MLP, and GRU, achieving accurate time-series forecasts with reduced computational and communication costs, showcasing FL's efficiency in large-scale environments.
Nevertheless, a further variation of FL is 
\emph{Reinforcement-learning based FL (FRL)}
which combines Reinforcement Learning principles with FL, training models across decentralised devices without centralised data sharing. An example of it is [S51] that introduces an FRL approach, combining it with the Asynchronous Advantage Actor-Critic (A3C) algorithm for Space-Air-Ground Integrated Network (SAGIN). This framework addresses challenges in resource management and transmission strategy at SAGIN.

\subsubsection{Reinforcement Learning (RL):} RL is the second most frequent learning type used in the studies, with 23.64\% reporting it, especially for JRO, Beam Selection, and optimisation. It can be further classified based on whether they are applying the Deep learning-based reinforcement learning method or optimisation-based RL.
\emph{Deep Reinforcement Learning (DRL)} is being used in the majority 76\% of the reinforcement learning approach. These approaches learn the policy to optimise the solution by training as a deep learning network. For example, [S7] proposed MATD3 (Multi-Agent Twin Delay Deep Deterministic Policy Gradient), a deep reinforcement learning technique for resource allocation and computation offloading in Cybertwin-enabled 6G wireless networks. MATD3 is designed to optimise quality of service by minimising latency and energy consumption, managing cache resources efficiently, reducing latency and energy consumption.

An alternative stream of RL is \emph{Optimisation-based RL}.
Two studies have employed optimisation-based Reinforcement Learning (RL), each addressing distinct aspects. The first study [S62] proposed the Context and Social-aware Machine Learning (CSML) approach, focusing on beam allocation using a two-layer Multi-Armed Bandit (MAB) algorithm for aggregated data in mmWave communication systems. CSML leverages context and social preference information for efficient beam coverage, employing a double-layer online learning algorithm with the MAB algorithm for beam and angle selection based on received data. 
In the second study [S72], MAMBA, a restless multi-armed bandit (MAB) framework for beam tracking and rate adaptation in 5G mmwave systems, was introduced. It employs the Adaptive Thompson Sampling (ATS) RL algorithm for optimal beam and coding scheme selection to maximize data rates in dynamic environments. The results show significant throughput gains and outperform traditional beam tracking algorithms.
\begin{table}[!tb]
\centering
\caption{Comparison of Dataset Types in AISM Research}
\label{tab:dataset_comparison}
\footnotesize
\resizebox{\textwidth}{!}{\begin{tabular}{|l|p{5cm}|p{5cm}|p{5cm}|}
\hline
\textbf{Type} & \textbf{Definition} & \textbf{Strengths} & \textbf{Weakness} \\
\hline
\multirow{6}{*}{Real} & Datasets derived from empirical observations or measurements in real-world environments. Examples include network traces from LTE networks, Channel Utilisation (CU) data from WLANs, and MIMO base station measurements. & - Provides insights into practical scenarios\leavevmode\newline - Facilitates realistic performance evaluation\leavevmode\newline - Useful in domains like Beam Selection, DSA, Traffic Prediction, Channel Estimation, and Spectrum Sensing & - Limited by the availability and accessibility of real-world data\leavevmode\newline - May contain privacy or security-sensitive information\leavevmode\newline - Can be outdated quickly due to the fast evolution of technology \\
\hline
\multirow{6}{*}{Simulated} & Datasets crafted through models or simulations to emulate real-world behaviors. Notable examples include the DeepMIMO dataset, datasets for SWIPT system CSI, and human activity recognition datasets. & - Can be generated in large quantities\leavevmode\newline - Allows for the modeling of hypothetical scenarios not yet observed\leavevmode\newline - Useful in a wide range of applications including JRO, DSA, Beam Selection, and Channel Estimation & - May not accurately represent real-world complexities\leavevmode\newline - Depends on the fidelity of the simulation models used\leavevmode\newline - Potential for overfitting to simulated scenarios that do not translate to real-world effectiveness \\
\hline
\multirow{6}{*}{Synthetic} & Artificially generated datasets designed to mimic certain characteristics of real data. Examples include datasets for Federated Learning scenarios, RadioML2016 for signal processing, and datasets with variations in modulation scenarios. & - Offers control over dataset parameters and conditions\leavevmode\newline - Enables testing against a wide range of scenarios and noise conditions\leavevmode\newline - Particularly useful in Beam Selection and Spectrum Sensing applications & - May lack the unpredictability and variability of real-world data\leavevmode\newline - Risk of not capturing all relevant real-world phenomena\leavevmode\newline - Can be challenging to generate synthetic datasets that are representative of complex real-world conditions \\
\hline
\end{tabular}}
\end{table}
\begin{table}[!tb]
\caption{Publicly available Datasets, Simulation Tools, and Testbeds used by AISM}
\footnotesize
\resizebox{\textwidth}{!}{%
\begin{tabular}{|>{\raggedright\arraybackslash}m{8mm}|m{12mm}|m{12mm}|m{15mm}|m{70mm}|m{10mm}|}
\hline
\textbf{AISM} &\textbf{Data Type} & \textbf{Dataset Type} & \textbf{Tool} & \textbf{Description} & \textbf{Study Ref}\\
\hline
\multirow{5}{*}{\rotatebox[origin=c]{90}{\parbox{1cm}{\textbf{Spectrum Sensing}}}} &\multirow{6}{*}{RF Signal} & Real & GNURadio v3.7.9. & Open-source toolkit for software-defined radio [\href{https://www.gnuradio.org/news/2016-07-05-gnu-radio-v3-7-10-release/}{GNURadio}]. & [S31] \\
\cline{4-6}
 & & & RFExplorer (WSUB1G) &Compact spectrum analyzer for sub-1GHz frequencies [\href{http://j3.rf-explorer.com/228}{RFExplorer}]. & [S29] \\
 \cline{4-6}
 & & & USRP2 testbed & Hardware setup using USRP for RF signal collection and analysis [\href{https://www.ettus.com/all-products/ub210-kit/}{USRP2 testbed}]. & [S36] \\
 \cline{3-6}
 & & Simulated & Electrosense & Crowd-sourced network for global spectrum data analysis [\href{https://electrosense.org/}{Electrosense}]. & [S49] \\
 \cline{3-6}
 & & Synthetic & RadioML 2016.10a & Dataset with samples across modulation schemes and SNR levels [\href{https://github.com/wzjialang/MCLDNN/tree/master}{RadioML 2016.10a}]. & [S67, S74, S76, S94, S95] \\
\hline
\multirow{8}{*}{\rotatebox[origin=c]{90}{\parbox{1.5cm}{\textbf{Beamforming Selection} }}}& Channel Data & Real + Simulated & 3D Raytracing Simulations & 3D ray-tracing for object position and distance analysis [\href{https://www.mathworks.com/help/vdynblks/ref/simulation3draytracer.html}{MathWorks}]. & [S66] \\
\cline{3-6}
 && \multirow{2}{*}{Simulated} & DeepMIMO & Customizable MIMO datasets via 3D ray-tracing [\href{https://www.deepmimo.net/}{DeepMIMO}]. & [S12, S67]
 \\
 \cline{4-6}
 & & & Wireless Insite & Suite for RF propagation and communication channel analysis [\href{https://www.remcom.com/wireless-insite-em-propagation-software/}{Wireless Insite}]. & [S9] \\
\cline{2-6}
&LIDAR Data & Simulated & Blender Sensor Simulation & 3D spatial data for autonomous driving and mmWave communication [\href{https://github.com/lasseufpa/5gm-lidar}{GitHub}]. & [S64, S96] \\
\hline

\multirow{3}{*}{\rotatebox[origin=c]{90}{\parbox{1.5cm}{
\textbf{Channel Estimation}
}}}& Channel Data & Real & Industrial dataset & RF data from industrial settings for propagation analysis [\href{https://www.nist.gov/ctl/smart-connected-systems-division/networked-control-systems-group/project-data-wireless-systems\#3}{NIST}]. & [S35] \\
\cline{3-6}
 & & Simulated & MATLAB’s 5G toolbox & Dataset with 256 sets for CNN training on channel parameters [\href{https://github.com/ocatak/6g-channel-estimation-dataset}{GitHub}]. & [S34] \\
 \cline{2-6}
 & RF Signal & Simulated and Synthetic &RADIOML 2018.01A & Dataset with synthetic, simulated channel effects for various modulation types [\href{http://deepsig.io/datase}{RADIOML 2018.01A}]. & [S6] \\
\hline

\multirow{4}{*}{\rotatebox[origin=c]{90}{
\textbf{DSA}
}}
 & RF Signal & Real & DARPA & Datasets for RF spectrum sharing innovations [\href{https://www.darpa.mil/program/spectrum-collaboration-challenge}{DARPA}]. & [S43] \\
\cline{3-6}
& & Simulated & LTE-V2V simulator & 
Simulator for LTE-V2V resource allocation, supporting IEEE 802.11p/ITS-G5 [\href{https://github.com/alessandrobazzi/LTEV2Vsim}{GitHub}]. & [S27] \\
\cline{2-6}
 & Channel Data& Simulated & Colosseum channel emulator & Emulator with 256 SDRs for AI and ML experiments [\href{https://www.northeastern.edu/colosseum/}{Colosseum}]. & [S39] \\
\hline
\multirow{2}{*}{\rotatebox[origin=c]{90}{\parbox{1cm}{
\textbf{JRO}
}}} & IMAGE DATA & Real & MNIST & Collection of 70,000 labeled images of handwritten digits [\href{https://www.kaggle.com/datasets/hojjatk/mnist-dataset}{Kaggle}]. & [S26] \\
\cline{2-6}
 & Channel Data & Simulated & THZ-Wireless-Channel & 
Channel impulse responses between 240GHz and 300GHz [\href{https://ieee-dataport.org/open-access/thz-wireless-channel-measurements-between-240ghz-and-300ghz}{IEEE DataPort}]. & [S78] \\
\hline
\rotatebox[origin=c]{90}{\parbox{1.5cm}{
\textbf{Traffic Predicition}
}}
& Network Traces & Real + Synthetic & Botnet Traffic Dataset & Offers labeled real-world botnet traffic across thirteen scenarios for cybersecurity research [\href{https://www.stratosphereips.org/datasets-ctu13}{StratosphereIPS}] & [S59] \\
\hline
\end{tabular}
}
\label{tab:dataset_table}
\end{table}
\begin{figure}[!tb]
\centerline{\includegraphics[width=\textwidth]{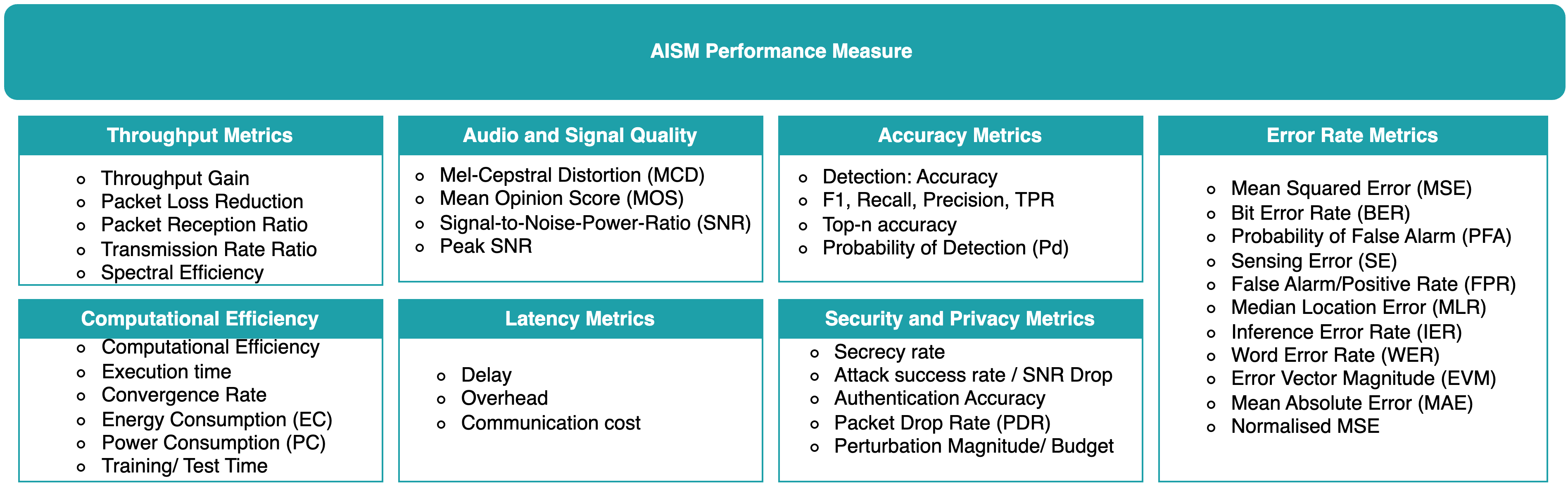}}
 \caption{Common Performance Measures used in AI-enabled Spectrum Management}
\label{Performance}
 \end{figure} 
 \begin{table}[!ht]
\centering
\footnotesize
\caption{Most Frequently Reported Performance Metrics and their Definitions}
\label{commonmetrics}
\begin{tabular}{|>{\raggedright\arraybackslash}m{30mm}|m{110mm}|}
\hline
\textbf{Metric} & \textbf{Definition}  \\
\hline
Accuracy & It is a measure of how well a system's outputs match the correct or expected outputs.   \\
\hline
Recall & It is the proportion of actual positive instances correctly identified by the model.  \\
\hline
Precision & It is the proportion of predicted positive instances that are actually positive.  \\
\hline
F1-score & F1-score is the harmonic mean of precision and recall, providing a balance between the two.  \\
\hline
Error Rate & Error Rate is a measure of the overall rate of misclassifications made by a model.  \\
\hline
Mean Square Error (MSE) & It measures the average squared difference between the actual values ($Y_i$) and the predicted values ($\hat{Y}_i$).   \\
\hline
Mean Absolute Error (MAE) & It measures the average absolute difference between the actual values ($Y_i$) and the predicted values ($\hat{Y}_i$).   \\
\hline
BER & Bit Error Rate (BER) measures the ratio of incorrectly received bits (BR) to the Total number of Transmitted Bits (TTB).   \\
\hline
Throughput & Throughput computes the rate of Successful data Transmission (ST) over a communication channel or network.   \\
\hline
Transmission Rate & It calculates the speed at which data is transmitted over a communication channel.   \\
\hline

Spectral Efficiency & Quantifies the efficiency of a communication system in utilizing limited frequency spectrum, measured in bits per second per hertz (bps/Hz).  \\
\hline
Execution Time & It measures the total time taken for a computational process to complete.  \\
\hline
ECR & Energy Consumption Rate (ECR) measures the ratio of consumed energy (CE) to the Total Energy Consumed (TEC).  \\
\hline
Authentication Accuracy & The probability of correctly identifying an illegitimate message (\(H_1\)).  \\
\hline
Secrecy Rate & The measure of secure information transmission capacity in a communication channel.  \\
\hline
Packet Drop Ratio (PDR) & The ratio of packets dropped to the total number of packets transmitted.  \\
\hline
MCD (Mel-Cepstral Distortion) & MCD is a metric used in speech and audio processing to measure the difference between the true and predicted mel-frequency cepstral coefficients (MFCCs).  \\
\hline
\end{tabular}
\end{table}

\subsection{RQ1.3: Dataset Taxonomy and Characteristics}
This section delineates the taxonomy and characteristics of datasets employed in studies within the domain of AISM. The datasets are broadly categorised into Real, Simulated, and Synthetic types. Their definitions, along with their strengths and weaknesses, are listed in Table~\ref{tab:dataset_comparison}. Moreover, Table~\ref{tab:dataset_table} meticulously lists publicly available datasets utilised in these studies, while Figure \ref{Overall_Stats_RQ1} (b) provides a comprehensive distribution overview across AISM applications.

\emph{Real Datasets:} About 23.6\% of the studies reviewed use real datasets, mainly in Beam Selection (27\%), Dynamic Spectrum Access (DSA) (15\%), Traffic Prediction (23\%), Channel Estimation, and Spectrum Sensing (12\% each). Notable examples include the MNIST dataset [S26], LTE network data from a Greek city [S2], Channel Utilization (CU) data from the University of Oulu [S3] and channel measurements for UAV task queues [S8]. For instance, [S26], use MNIST dataset to simulate their proposed FL algorithm for minimizing Energy Consumption (EC), optimizing the device pairing, communication resource and computation resource.
\emph{Simulated Datasets:} Over 51.8\% of the research utilizes simulated datasets, especially in Joint Resource Optimization (JRO) (26\%), DSA (20\%), Beam Selection (17\%), and Channel Estimation (11\%). Key examples include the DeepMIMO dataset [S47, S59] and a SWIPT system CSI dataset [S34].
\emph{Synthetic Datasets:} Constituting 14.5\% of studies, synthetic datasets are mainly used in Beam Selection (31\%) and Spectrum Sensing (23\%). Examples include a Federated Learning scenario by [S51] and the RadioML2016 dataset used in [S67, S74, S76] for performance evaluation across various signal conditions. Another example is a dataset offering diverse modulation scenarios [S40].

\subsection{RQ1.4: Performance Evaluation Metrics}
\vspace{1em}
This section provides an in-depth exploration of the performance metrics employed by AISM. Figure \ref{Performance} presents a taxonomy and prevalent performance measures utilised in the reviewed studies. Upon analyzing the distribution of these metrics, it becomes apparent that error rate and accuracy metrics are the most frequently reported indicators, comprising 22.58\% and 21.51\%, respectively. These metrics play a fundamental role in assessing the precision and correctness of diverse systems or models. Additionally, Throughput and computational efficiency are noteworthy, representing 20.43\% and 17.20\% of the metrics, respectively. Throughput reflects the overall capacity to transmit data, while computational efficiency is critical for evaluating resource utilisation in algorithms or systems. The definitions for the most common metrics reported by the studies are enlisted in Table ~\ref{commonmetrics}.
Figure \ref{Overall_Stats_RQ1} (c) presents the adaption of these metrics across the AISM solutions. It can be seen that
\textit{Error rate} emerges as the most common metric (20\%) in the reviewed studies. Notably, it has a frequent application in Channel Estimation, Traffic Prediction, and JRO studies.
\textit{Accuracy metrics}, comprising various measures such as Detection accuracy, F1score, Recall, Precision, TPR, Top-n accuracy, and probability of detection, are the second most popular (19.09\%) among the reviewed studies. These metrics are frequently reported in studies related to DSA, Traffic Prediction, Spectrum Sensing, and Beam Selection.
Throughput metrics are prominently featured, constituting a significant 18.18\% of reported metrics. The emphasis on throughput is particularly apparent in studies related to Beam Management.
Computational efficiency, measuring algorithmic or computational system effectiveness within a given time frame, is reported in more than 15.45\% of the studies. This metric is commonly addressed in studies related to JRO, Beam optimisation, Beam Selection, and Spectrum Sensing.
Security and privacy metrics are reported in 12.73\% of the studies, prominently in Spectrum Sensing and JRO applications. Authentication accuracy, secrecy rate, and Packet Drop Ratio (PDR) are among the metrics employed to evaluate security aspects. 
Latency metrics, crucial in the context of future 6G requirements for low-latency and real-time environments, are reported in 10.91\% of the studies. These metrics include delay, overhead, and communication cost, particularly in JRO applications. 
Audio and signal quality metrics are used in 3.64\% of the reviewed studies, primarily in the context of beam selection. Examples include studies evaluating acoustic eavesdropping methods [S20] and signal quality improvement using measures such as Mel-Cepstral Distortion (MCD) and Peak Signal-to-Noise Ratio (PSNR) [S38]. 

\begin{small}
\noindent\fbox{%
\parbox{\textwidth}{%
 \textbf{\textit{Summary}}: 
\begin{itemize}
\item Our SLR revealed that AISM applications span Resource, Beam, and Channel Management, with a notable emphasis on AI methodologies in dynamic allocation, traffic prediction, and joint resource optimization in resource management (39.78\% of studies), adaptive beam selection, alignment, optimization, and tracking in beam management (29.09\%), and advanced channel estimation and spectrum sensing techniques in channel management (23.67\%).
\item  The taxonomy of learning approaches in AISM emphasizes the prevalence of Deep Learning (DL) and Reinforcement Learning (RL), with DL primarily utilized for beam selection and DSA, and RL frequently applied in JRO and beam optimization.
\item Over 66\% of studies use simulated or synthetic datasets. 
 \item Error rate and accuracy are the predominant performance metrics in AISM.
\end{itemize}}}\end{small}

\section{RQ2: Security and Privacy (S\&P) Challenges}
\label{sec:securityandprivacy}

\begin{table*}[!tb]
\centering
\caption{RQ2.1: Classification of S\&P  Threats on B5G Spectrum Management systems}
\footnotesize
\resizebox{\textwidth}{!}{\begin{tabular}{|p{0.75cm}|p{2cm}|p{3cm}|p{6cm}|p{3cm}|}
\hline
\textbf{Type}               & \textbf{Threat}                & \textbf{Sub-Type}             & \textbf{Definition}                                                                                         & \textbf{Study Mapping} \\ \hline
\multirow{12}{*}{\rotatebox[origin=c]{90}{\textbf{Adversarial}}}

& \multirow{6}{*}{Poisoning }            & Data Falsification     & Altering or injecting deceptive data into a system to mislead decision-making or corrupt AI models. & {[}S31, S32, S49, S97{]} \\ \cline{3-5}
& &Backdoor Attacks      &  Embed hidden triggers in a model to cause specific misbehaviors, unlike normal poisoning that broadly degrades performance. & {[}S96, S98{]} \\ \cline{2-5}
&\multirow{2}{*}{Evasion}             & Adversarial Example Generation & Generating or defending against manipulated inputs to deceive AI models.                        & {[}S6, S12, S18, S24, S40, S56, S94, S95{]} \\ \cline{2-5}
&\multirow{3}{*}{Inference}  & Membership Inference Attacks    & Aim to determine if specific data was used in a model's training, threatening privacy by exploiting model vulnerabilities. & {[}S33, S45{]} \\ \hline

\multirow{22}{*}{\rotatebox[origin=c]{90}{\textbf{Inherited}}}         
& \multirow{10}{*}{Eavesdropping }          & Acoustic Eavesdropping     & Using sound waves to intercept communications or infer information.                               & {[}S20, S85{]} \\ \cline{3-5}
                    && Transmission Channel Eavesdropping & Unauthorized access to data transmitted across communication channels.                             & {[}S51, S101, S103, S106{]} \\ \cline{3-5}
                    & &Localization and Activity Monitoring & Using signal analysis or environmental sensing to infer location or activities.                  & {[}S86, S87{]} \\ \cline{3-5}
                    & &Energy-Harvesting Eavesdropping  & Attacks leveraging wireless energy transfer or harvesting to intercept information.                & {[}S34, S50, S53{]} \\ \cline{3-5}
                    & &Passive Eavesdropping          & Stealthy attacks where eavesdroppers silently capture data without altering the transmission.     & {[}S8, S11, S14, S92{]} \\ \cline{2-5}
                    & \multirow{10}{*}{Spoofing }                             & RF Fingerprinting Spoofing   & Attacks that mimic the unique RF characteristics of legitimate devices to gain unauthorized access or privileges. & {[}S4{]} \\ \cline{3-5}
                    &                                       & Pilot Spoofing Attack (PSA) & A security threat where an eavesdropper transmits identical pilot sequences as a legitimate user to compromise data transmission. & {[}S105, S107{]} \\ \cline{3-5}
                    &                                        & Identity Spoofing in mmWave Networks & Attacks that impersonate a legitimate user or device in millimeter wave (mmWave) networks by exploiting vulnerabilities in the network's identity verification processes. & {[}S108, S109, S110{]} \\ \cline{2-5}
                    & \multirow{12}{*}{Jamming}                              & Intelligent Jamming          & Jamming attacks that use AI or ML to dynamically change their strategies for optimal disruption, considering the ongoing communication patterns and network defenses. & {[}S5, S17, S46, S47, S110{]} \\ \cline{3-5}
                    &                                      &   Reactive Jamming             & Attacks that are initiated in direct response to detected communications, aiming to disrupt these activities through immediate jamming upon sensing transmissions. & {[}S48{]} \\ \cline{3-5}
                    &                                       &  Operational Jamming          & Attacks targeting the specific vulnerabilities inherent to certain communication technologies, network architectures, or operational contexts, like 5G, MEC networks, or V2X communications. & {[}S27, S100, S102{]} \\ \cline{2-5}
                    & Data Leakage                          & User and Auction Data Leakage & Unauthorized disclosure of sensitive information due to insufficient data protection, distinct from direct communication interception.    & {[} S23, S25, S99, S104{]} \\ \hline
\end{tabular}}

\label{tab:RQ2_wireless_attacks}
\end{table*}

This section presents a concise overview of our thematic analysis based on items D17 to D22 from the Data Extraction Form, as shown in Table \ref{dataextractionform}. Figure \ref{fig:num_reference}(a) showcases the distribution of studies focusing on security and privacy within our review. Notably, 53\% of the studies we examined address security and privacy issues related to spectrum management systems, which underscores the importance of security as a critical issue in B5G networks.

\subsection{Classification of Threats on B5G and Future Generation Spectrum Management}
\begin{figure}[!tb]
\centerline{\includegraphics[width=\textwidth]{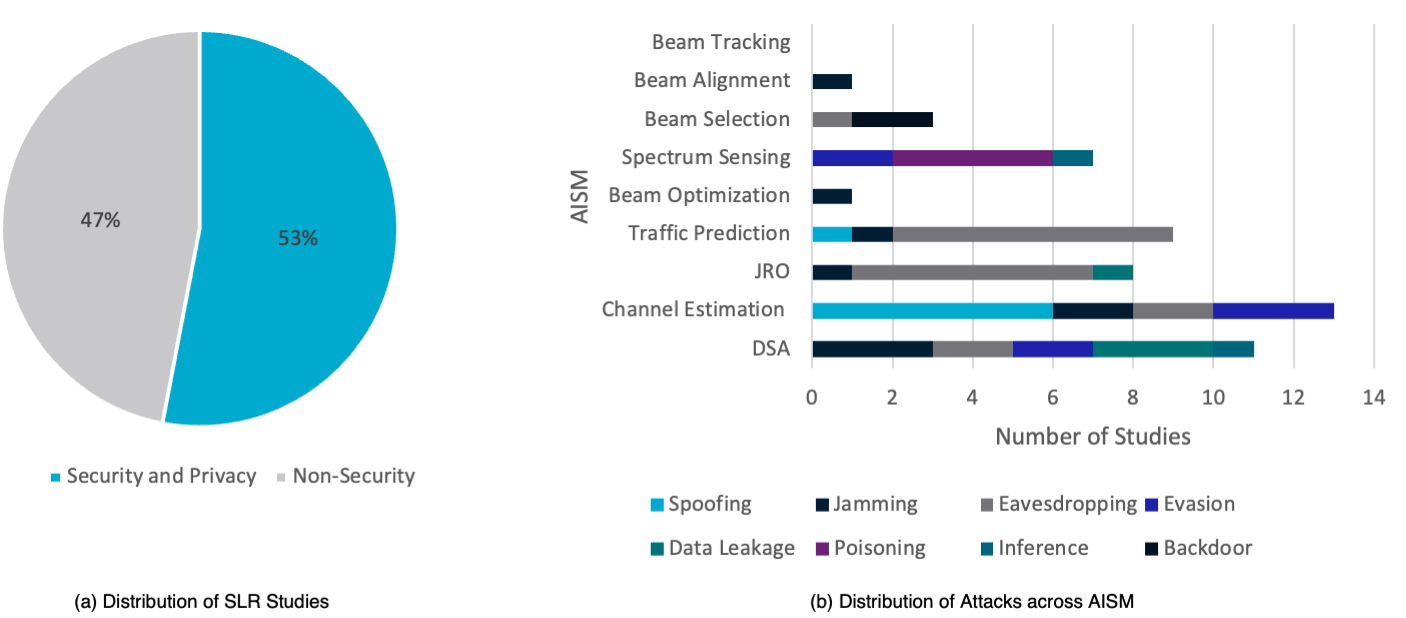}}
\caption{Distribution of Security and Privacy studies}
\label{fig:num_reference}
\end{figure}

Table \ref{tab:RQ2_wireless_attacks}, provides a taxonomy, its definition and study mapping for the attacks against Spectrum Management systems for B5G networks. Moreover, Figure \ref{fig:num_reference}(b) provides the distribution of AISM based on the attack classification. The attacks are mainly divided into two main categories, adversarial and inherited as discussed in the subsequent sections.
 
\subsubsection{Adversarial}
Adversarial Attacks are deliberate manipulations targeting AI-driven components, aiming to deceive or compromise their decision-making processes. 30.8\% of the security-related studies focused on adversarial attacks. These studies can be sub-classified into three main attack types: Poisoning, Evasion, and Inference. 

Among these studies, 50\% of them focus on \textbf{evasion} attacks. An evasion attack is a strategy employed at test time, where specifically crafted input data is used to deceive a machine learning model into making incorrect predictions or classifications.
For instance, [S6] demonstrates how crafting specific inputs can trick deep learning systems in applications like radio fingerprinting and modulation classification. 
Similarly, [S18] introduce RAdio Frequency Attack (RAFA), an evasion attack framework targeting FIRE, an ML method for forecasting downlink channels in FDD MIMO communication systems without feedback from client devices, and DLoc, a deep learning algorithm for precise indoor localization capable of overcoming multipath and occlusion issues. RAFA's primary aim is to introduce adversarial perturbations into the wireless channel estimation process, which is vital for the functionality of both FIRE and DLoc, thereby degrading MIMO efficiency and amplifying localization inaccuracies. This is accomplished by calculating and transmitting perturbations that interfere with channel estimation, utilizing the Projected Gradient Descent (PGD) method to refine the attacks.
Additionally, studies [S12, S24, and S56] delve into how AI-enabled applications and models in NextG networks are prone to adversarial evasion attacks, encompassing areas such as spectrum sensing and channel estimation. For instance, [S56] employs a variety of techniques—Fast Gradient Sign Method (FGSM), Basic Iterative Method (BIM), Momentum Iterative Method (MIM), Projected Gradient Descent (PGD), and Carlini \& Wagner (C\&W)—to create adversarial examples. These examples are then used to test the resilience of CNN-based channel estimation models against such attacks.

The \textbf{poisoning} attack, identified in 37\% of the adversarial attack studies, it targets the training data of ML models, leading to compromised decision-making and system functionality from within.
Notably, studies such as [S31, S32, and S49] delve into Spectrum Sensing \emph{Data Falsification} (SSDF) attacks, demonstrating sophisticated strategies to undermine spectrum sensing models' integrity. For instance, [S31] introduce a Learning-Evaluation-Beating (LEB) framework that successfully manipulates fusion center decisions by altering sensed data, while [S32] highlights stealthy, energy-efficient attacks to deceive transmitter spectrum sensing, affecting both operational and training phases. Additionally, [S97] investigates data poisoning in dynamic crowd-sourcing for spectrum sensing. On the other hand, studies [S96 and S98] explore another variant of poisoning attacks, i.e.,  \emph{backdoor}. For example, [S96] describes a stealthy backdoor attack on neural network-based beam models via data poisoning, embedding triggers that cause specific errors without general performance degradation, countered by a novel machine unlearning technique.
Furthermore, [S98] explores backdoor attacks in federated learning (FL) mmWave beam selection systems, introducing location-based triggers to compromise beam predictions and challenging traditional defenses.

\textbf{Inference attacks}, although less prevalent in the literature, present a unique threat by attempting to extract sensitive information from models. Two studies [S33,S45] examine how such attacks can reveal critical data. For instance, [S33] investigates the susceptibility of machine learning-based spectrum sensing models to membership inference attacks, demonstrating how these attacks can unveil sensitive data about signal characteristics. Whereas [S45] considers DSA systems, shows the vulnerability posed by inference attacks and proposes a method to optimize the delicate balance between user privacy and spectral efficiency.
\subsubsection{Inherited Attacks}
Inherited attacks exploit vulnerabilities in the underlying network infrastructure, many of which are carried over from previous generations of wireless networks. These attacks do not necessarily target AI components but rather the foundational aspects of the network that remain vulnerable. 69.2\% of the security-related studies focused on four types of inherited attacks: eavesdropping, Spoofing, Jamming, and Data Leakage. Approximately 40.54 of studies addressing inherited attacks consider \textbf{eavesdropping} as a significant threat model, where an unauthorized individual intercepts communications between two parties without their consent. These studies are categorized into five distinct sub-types of eavesdropping attacks, as detailed in Table \ref{tab:RQ2_wireless_attacks}. 
The first category \emph{acoustic eavesdropping} is exemplified by studies [S20] and [S85]. These studies employ mmwave radar technology to capture conversations through the detection of sound-induced vibrations. Notably, mmEcho [S20] achieves non-invasive audio capture through walls by sensing minor surface vibrations caused by sound waves. Similarly, mmSpy [S85] illustrates the capability of radar devices to reconstruct audio from phone conversations, highlighting significant privacy concerns related to acoustic vibrations.
\emph{Transmission channel eavesdropping} represents another facet of eavesdropping, characterized by unauthorized data access during transmission across communication channels. Investigations such as [S51] and [S101] propose advanced data security strategies. Specifically, [S51] explores the application of federated learning for decentralized data processing to mitigate eavesdropping risks within integrated networks. Concurrently, [S101] introduces a dual-layered defense mechanism for device-to-device communication, merging covert communication with friendly jamming techniques to safeguard against eavesdroppers, thereby illustrating the complexities involved in securing contemporary communication infrastructures.
The \emph{localization and activity monitoring} category signifies an interesting aspect of eavesdropping, focusing on the deduction of location or activities through signal analysis. [S86] presents mmTrack, leveraging 60GHz mmwave radios for precise indoor localization and tracking, thereby showcasing the potential of radio frequency signals for accurate user tracking. Likewise, [S87] proposes the m-Activity system, utilizing mmWave sensing for real-time activity recognition, emphasizing the utility of signal analysis in accurately monitoring human-oriented movements.
\emph{Energy-harvesting eavesdropping} encompasses attacks that exploit wireless energy transfer systems for information interception. Research such as [S34] and [S53] addresses the challenge of securing transmissions in scenarios where energy is concurrently transferred wirelessly alongside data. [S34], for instance, investigates the use of deep learning for resource allocation to combat eavesdropping in energy transfer networks, while [S53] employs full-duplex communication and artificial noise to enhance security against eavesdropping threats in industrial environments.
Finally, \emph{passive eavesdropping} involves the covert acquisition of data without altering the transmission. Efforts to counteract such threats include [S8] and [S11], which focus on enhancing secure computation and developing secure system architectures. [S8] emphasizes UAV-assisted Mobile Edge Computing, and [S11] addresses the security of non-orthogonal multiple access-based systems, both exemplifying proactive measures to bolster defenses against secret surveillance activities.

16\% of the studies in inherited attack studies considered \textbf{spoofing} as a threat model, where an adversary masquerades as a legitimate entity to deceive systems or individuals. These studies can be divided into three sub-types of spoofing attacks.
The first category, \emph{RF Fingerprinting Spoofing}, is exemplified by a study [S4], which presents an advanced deep learning method for analyzing transmitter fingerprints to enhance the security of critical infrastructures against unauthorized access. This approach significantly improves RF identification accuracy, showcasing its importance in countering spoofing attacks that mimic legitimate devices' RF characteristics.
The \emph{pilot Spoofing attack (PSA)} category, highlighted by studies [S105] and [S107], addresses PSA in channel estimation, where attackers mimic legitimate pilot sequences to disrupt data transmission. [S105] employs the Spatial Spectrum Method for MIMO system protection, and [S107] introduces enhanced training sequences with secure beam-forming to counteract PSA, emphasizing the need for robust defenses to ensure secure wireless communication.
\emph{Identity Spoofing:} It encompasses studies [S108, S109, S110], focusing on the detection and prevention of identity spoofing in the context of 5G mmWave networks. For instance, [S108] employs Principal Components of Channel Virtual Representation for accurate spoofing detection, whereas [S109] leverages unique SNR traces during the sector level sweep process for identifying spoofing attacks against IEEE 802.11ad devices. Interestingly, [S110] explores a novel scenario involving a full-duplex multi-antenna eavesdropper performing concurrent spoofing and jamming attacks, highlighting the complexity of safeguarding massive MIMO systems against sophisticated spoofing tactics. 

In spectrum management for Beyond 5G (B5G) networks, approximately 24.32\% of studies investigating inherited threats have pinpointed \textbf{jamming attacks} as a critical concern. These attacks, characterized by their intent to disrupt communication by flooding the frequency spectrum, are studied across three distinct types (see Table \ref{tab:RQ2_wireless_attacks}).
Firstly, \emph{intelligent jamming}, as explored in five key studies, leverages advanced machine learning algorithms to combat sophisticated jamming strategies. For example, research encapsulated in [S5] introduces a multi-user game model augmented with deep reinforcement learning to thwart adaptive jammers in Mobile Edge Computing (MEC) networks. Concurrently, [S17] utilizes federated deep reinforcement learning to address the challenges posed by smart jamming in two-tier 5G Heterogeneous Networks (HetNets), with a particular focus on optimization challenges such as beamforming and power allocation. 
The second type, \emph{reactive jamming}, receives specific attention in [S48] through the deployment of the "DeepFake" strategy. This method involves broadcasting deceptive signals designed to disorient and neutralize the efforts of reactive jammers.
The category of \emph{operational jamming}, is discussed in studies [S27, S100, S102], which concentrate on shielding crucial communication protocols against jamming interference. For instance, [S27] proposes a feedback-based attack detection mechanism combined with a fuzzy inference system for resource reservation as a countermeasure to packet dropping attacks in New Radio Vehicle-to-Everything (NR-V2X) communications. Similarly, [S102] highlights the vulnerability of beam training protocols in 5G millimeter-wave (mmWave) communication systems to jamming attacks, underlining the essential need to secure operational dimensions of wireless networks.
Lastly, \textbf{Data leakage}, emerges as a critical privacy concern in reviewed studies, where the unauthorized access or disclosure of sensitive information poses risks. In [S23], the threat is linked to federated learning processes, highlighting how data aggregation or training phases could expose sensitive information without adequate safeguards. Moreover, [S25] examines data leakage during the proactive caching's data analysis phase, pointing to potential mishandling or unauthorized access to user data, such as location and preferences. Meanwhile, [S99 and S104] explore the threat in the context of auction systems, where the leakage of bidders' sensitive information, including geo-locations and bid values, could occur through the platform, emphasizing the need for robust privacy protections across different phases of data handling and processing.
\subsection{RQ2.2: Classification of Defenses against S\&P Threats}
Table \ref{table:RQ2_cybersecurity_defenses} presents a systematic categorization of defense mechanisms, providing definitions and correlating them with the studies that tackle the security and privacy (S\&P) threats identified earlier. This table organizes these defense strategies into four main categories, each addressing one or more types of attacks. 

\begin{table*}[!tb]
\centering
\caption{Overview of Defenses and Their Targeted Attacks}
\label{table:RQ2_cybersecurity_defenses}
\footnotesize
\resizebox{\textwidth}{!}{\begin{tabular}{|p{2cm}|p{2cm}|p{7cm}|p{4cm}|}
\hline
\textbf{Defenses} & \textbf{Sub-Type}& \textbf{Their Definition} & \textbf{Attack [Studies Mapped]} \\
\hline
\multirow{3}{*}{\textbf{{\parbox{1.75cm}{Anomaly Detection}}}}&Behavioral& Focus on identifying deviations from established "normal" behavior patterns within specific contexts, by detecting significant deviations from these baselines. & Poisoning [S49], Eavesdropping [S28, S59], Spoofing [S108, S109], \\ 
\cline{2-4}
& Outlier Detection& Employs statistical analysis and data correlations to detect anomalies, targeting malicious activities or system faults in complex data environments.&Poisoning [S31, S97], Eavesdropping [S54, S60]\\
\hline
\multirow{6}{*}{\textbf{{\parbox{1.65cm}{Model Hardening}}}} & Adversarial Training &Process of training models with adversarial examples to enhance their defenses against attacks. & Poisoning [S96], Evasion [S18], Jamming [S32]\\ \cline{2-4}
&Noise or Smoothing& Boost model robustness through noise addition or smoothing techniques.& Poisoning [S98], Evasion [S40], Inference [S33], Eavesdropping [S85] \\ \cline{2-4}
&Defensive Distillation & Enhances model resilience by training a second model on softened outputs from a first model to smooth decision boundaries against adversarial inputs. & Evasion [S12, S24, S56]  \\
\hline
\multirow{8}{*}{\textbf{{\parbox{1.65cm}{Privacy Preserving Technique}}}}&Federated Learning & Builds global models without central data collection, preserving privacy by processing sensitive information locally. & Inference [S23], Eavesdropping [S51], Data-leakage [S25], Jamming [S17] \\ \cline{2-4}
&Decentralized Spectrum Privacy & Utilizes decentralized strategies and technologies like cognitive radio and blockchain for spectrum sharing while ensuring user privacy. & Inference  [S45], Eavesdropping [S103], Data leakage [S104] \\  \cline{2-4}
&Data Privacy Enhancements & Protects data using cryptographic methods, secret sharing, and obfuscation to secure data during processing and sharing. & Data leakage [S99, S104] \\  \cline{2-4}
&Differential Privacy & Applies differential privacy to anonymize data or query results, safeguarding individual data privacy against inference attacks. & Eavesdropping [S50]\\
\hline
\multirow{8}{*}{\textbf{{\parbox{1.65cm}{Physical Layer Security}}}}
&Secure Transmission and Signal Integrity &Strategies, architectures, and techniques focusing on securing communication channels, optimizing secrecy rates, maintaining signal integrity, and addressing security in next-gen networks.  & Eavesdropping [S11, S14, S20, S34, S92, S101, S106, S105, S107]\\
\cline{2-4}
&Anti-Jamming Strategies & Designed to protect spectrum management systems from intentional interference or jamming, often leveraging advanced algorithms to adaptively counteract or mitigate jamming effects.	&Jamming [S5, S17, S47, S48, S100, S102]\\ \cline{2-4}
&RF Authentication and Analysis & Techniques leveraging RF signal characteristics and physical channel properties for device identification and authentication, including advanced channel analysis for security. & Spoofing [S4, S35] \\ 

\hline
\end{tabular}}
\end{table*}

In our reviewed studies \textbf{Anomaly Detection}, emerges as a key defensive strategy. It utilizes machine learning or statistical techniques to detect and adapt to novel attacks or secure against anomalies in network communications by identifying anomalous behavior or data patterns. It is featured prominently in nine distinct primary studies that can be classified into behavioral anomaly and outlier detection as defined in Table \ref{table:RQ2_cybersecurity_defenses}.
Five studies delve into \textit{behavioral} anomaly detection, showcasing novel methods to protect wireless networks from various cyber threats by analyzing complex patterns and anomalies in network behavior. For example, [S28] introduces a cognitive network system architecture designed for environments riddled with trust challenges. It employs a fusion of multi-feature optimization algorithms and deep reinforcement learning to improve spectrum prediction, utilization, and sensing accuracy, efficiently pinpointing untrustworthy users. Similarly, [S49] introduces CyberSpec, a framework that combines ML and DL to detect Spectrum Sensing Data Falsification (SSDF) attacks through device behavioral fingerprinting, showcasing its effectiveness on the ElectroSense platform.
In contrast, [S59] outlines a cyber defense architecture specifically designed for 5G networks, employing deep learning to analyze network traffic. This system uniquely adapts its configuration to traffic fluctuations, optimizing detection processes to maintain high efficiency against anomalies. 
Likewise, [S108] proposes a novel mechanism to combat identity spoofing in 5G networks, especially focusing on mmWave massive MIMO systems. It uses a Neyman-Pearson (NP) testing-based approach and a single-hidden layer feedforward neural network for dynamic scenarios, relying on Principal Components of Channel Virtual Representation (PC-CVR) to achieve high detection rates.
On the other hand, \textit{outlier detection}, is employed in four studies to identify anomalies. [S31] uses an Influence-Limiting Defence strategy to explore protecting cooperative spectrum sensing from Low-Effort Jamming (LEJ) attacks. Similarly, [S97] delves into countering malicious data poisoning in crowd-sourcing through online quality learning. In contrast, [S54] presents a method focusing on anomaly detection in the Industrial Internet of Things (IIoT), specifically for safeguarding against eavesdropping attacks, by introducing a multidimensional data relationship algorithm. Likewise, [S60] proposes a technique for anomaly detection in urban sensing that differentiates between genuine errors and significant events, highlighting the role of outlier detection in pinpointing anomalies.

\textbf{Model Hardening:} It emerges as a dominant defensive strategy against a variety of adversarial attacks highlighted in this SLR, finding its application in ten studies. It involves modifications to the training process or model parameters, including retraining or direct modifications to model architecture, to mitigate effects of attacks or improve resilience against adversarial attacks. 
This approach is categorized into three distinct types of model-hardening defenses:
\textit{Adversarial Training:} 
[S18] utilizes adversarial training to increase the resilience of FIRE against Universal Adversarial Perturbations (UAP) by incorporating batch-wise perturbations through the RW-PGD algorithm, significantly boosting the model's robustness against RAFA attacks, albeit at the cost of extended training time. On the other hand, [S32] introduces a form of real-time adversarial training into the communication process. By intentionally transmitting incorrect signals at strategic times, the defense mechanism is effectively generating adversarial examples in the operational environment. These actions prevent the adversary from accurately learning the system's normal operating behavior, which is a direct form of adversarial training. Unlike traditional adversarial training, which is done during the model training phase to improve its resilience, this strategy applies the concept dynamically, in real-time, to disrupt the adversary's learning process.
\textit{Noise or smoothing Strategies:} These are reported in three studies [S33] that employ randomized smoothing to enhance the resilience of wireless signal classification by augmenting the training dataset with Gaussian noise, thereby countering adversarial perturbations. [S40] presents a defense mechanism for automatic modulation classification (AMC) models in spectrum sensing by integrating Spatial Transformer Networks (STN) with a novel ResNeXt architecture, termed STN-ResNeXt, and applying transfer learning (TL) to adjust to new signal conditions. Lastly, [S98] introduces a dynamic norm clipping and federated noise titration approach to detect and counter backdoor attacks in federated learning (FL) systems, securing FL environments from backdoor threats while maintaining high accuracy for legitimate tasks.
\textit{Defensive Distillation:} Three studies used defensive distillation as a defense against evasion attacks. [S12] investigates the implementation of AI-powered Intelligent Reflecting Surfaces (IRS) and recommends defensive distillation as a method to enhance robustness.
Lastly, [S56] offers a detailed examination of deep learning (DL)-based channel estimation models, showing how defensive distillation effectively protects these models from adversarial manipulations.

\textbf{Privacy Preserving Techniques:} These techniques are aimed at safeguarding user privacy by minimizing exposure to potential inference attacks, balancing the trade-off between preserving privacy and maintaining system utility. Eight studies used privacy-preserving methods to protect the integrity of data from inference, eavesdropping, and data leakage attacks used in spectrum management. These studies are further sub-classified into four types.
\textit{Federated Learning (FL):} FL is prominently featured in studies [S23, S25 and S51]. These studies highlight FL's capacity to build global models without central data collection, thus preserving privacy by processing sensitive information locally. For instance, [S23] introduces FL into a two-level Multi-access Edge Computing (MEC)-assisted vehicular network framework to address privacy concerns in Intelligent Transportation Systems (ITS). 
Whereas [S51], explores the integration of federated learning (FL) into Space-air-ground Integrated Networks (SAGIN) to manage resource and transmission strategies securely and efficiently. FL's decentralized nature provides privacy and security guarantees, reducing the risk of eavesdropping attacks.
\textit{Decentralized Spectrum Privacy:} It is the focus of studies [S45, S103, and S104]. For example, [S45] develops a privacy-preserving spectrum-sharing system architecture that addresses the privacy-performance tradeoff. It employs optimal and heuristic strategies to maximize privacy while minimizing the impact on spectrum efficiency. 
\textit{Data Privacy Enhancements:} Studies [S25, S99, S104], propose these methods to protect the privacy of the data. For example, [S25] introduces a federated K-means scheme for proactive caching in cellular networks, integrating secret sharing to protect user data during clustering. [S99] introduces a novel framework for enhancing privacy and efficiency in spectrum auctions. It employs lightweight cryptographic techniques and a data-oblivious auction algorithm to secure bidders' privacy without compromising spectrum allocation efficiency. 
\textit{Differential Data Privacy:} It is applied in one study [S50], which presents a secure and intelligent energy harvesting framework for privacy-preserving 6G-enabled IoT. It integrates differential privacy (DP) and intelligent reflecting surface technologies to protect users' locations from eavesdropping attacks while optimizing the energy harvesting process.

\textbf{Physical Layer Security:} Encompasses strategies and techniques leveraging the physical characteristics of communication channels to enhance data protection against eavesdropping, jamming, and spoofing, using dynamic protocols, signal integrity optimization, AI, and RF analysis. 43.39\% of the security-related studies cover this defense type. These studies can be sub-classified as enlisted in Table \ref{table:RQ2_cybersecurity_defenses}.
\textit{Secure Transmission and Signal Integrity:}
Nine out of twenty-three studies under physical layer security are categorized under Secure Transmission and Signal Integrity. 
For instance, [S11] presents a secure system design for NOMA-based MEC systems, incorporating hybrid SIC decoding to effectively counter eavesdropping attacks, optimizing both latency and task processing. In contrast, [S14] introduces a deep learning-based secure transceiver design for 5G networks using a VAE, which ensures minimal data leakage even with imperfect CSI, significantly enhancing communication quality and security. 
Whereas [S34] employs AI, specifically DNN methodologies, to optimize power allocations in SWIPT networks, achieving near-optimal secrecy rates with considerably reduced computational demands. 
\textit{Anti-Jamming Strategies:}
Numerous innovative anti-jamming strategies for beyond 5G networks have been examined in our SLR. For example, [S17] proposes a federated deep reinforcement learning technique for defending against jamming in 5G heterogeneous networks. This method optimizes beamforming vectors and power allocation, enhancing the rate for femto users despite jamming, marking a departure from traditional strategies like frequency hopping by leveraging federated learning and Dueling Double Deep Q Networks for adaptive and collaborative learning. In contrast, [S47] introduces a novel defense against smart and reactive jamming using deep reinforcement learning and ambient backscattering communications, achieving substantial improvements in throughput and packet loss reduction by dynamically adapting to the jammer's strategy without prior knowledge. 
Another example is study [S102], it proposes a two-stage machine learning strategy for protecting mmWave communications in smart factories, detecting and mitigating jamming attacks to improve beam alignment decisions, and showcasing significant enhancements in communication reliability and security. These studies collectively showcase progress in securing wireless networks against jamming threats through advanced anti-jamming strategies, including ML-based detection, mitigation, strategic deception, and adaptive learning.

\textit{RF Authentication and Analysis:}
Within the application of channel management, RF Authentication and Analysis have been found in two investigations. The first, [S4], employs a deep learning method with data augmentation for RF fingerprinting, achieving 97.84\% accuracy in authenticating transmitter signals, significantly outperforming previous models. The second, [S35], introduces a ML-based, threshold-independent authentication technique using MIMO technology and channel state information in industrial wireless networks, offering higher accuracy and optimization potential than traditional methods. Together, these approaches mark a significant leap in RF security, highlighting the importance of machine learning in enhancing authentication and protecting against cyber threats.

\begin{small}
\noindent\fbox{%
\parbox{\textwidth}{%
 \textbf{\textit{Summary}}: 
  \begin{itemize}
      \item Our SLR has unveiled that security and privacy considerations are of paramount importance in B5G networks, with over half of the studies examined emphasizing these issues within spectrum management systems.
      \item Thematic analysis highlights the emergence of adversarial attacks, such as evasion, poisoning, and inference, alongside traditional threats like eavesdropping, spoofing, jamming, and data leakage, as significant vulnerabilities exploiting network infrastructure and AI components.
      \item Identified countermeasures in the literature include anomaly detection for threat identification, model hardening to bolster attack resilience, privacy-preserving mechanisms to safeguard user data, and physical layer security to counteract eavesdropping and jamming. 
     
  \end{itemize}}}
  \end{small}
\section{Discussion, Challenges and Future Direction}





In this section, we summarize our most important findings and discuss promising research directions that may help further the field. 


\subsection{Under-addressed AI usage}
Our analysis in RQ1 (Section \ref{section_RQ1} reveals a notable gap in the exploration of AI-enabled applications within Beam Alignment and Beam Tracking, comprising only 2.72\% of studies, despite their critical importance in ensuring efficient and reliable communication in high-frequency networks, particularly in the context of 6G. AI presents a promising avenue for addressing the dynamic challenges of beam management, offering precision, adaptability, and real-time decision-making capabilities \citep{guan2019efficiency}. To fully harness the potential of AI in enhancing beam management for 6G networks, future research efforts should prioritize the development of specialized AI models and algorithms tailored to the unique requirements of high-frequency communication systems. This necessitates interdisciplinary collaboration between the AI and telecommunications research communities to bridge the current research gap and pave the way for transformative advancements in wireless communication technologies.
 \subsection{Computational resource requirements}
In RQ1, DL and RL emerge as the predominant methodologies utilized in 47.27\% and 23.64\% of AI-enabled Spectrum Management (AISM) studies, respectively. These techniques, applied across diverse applications such as beam selection, DSA, traffic prediction, and JRO, showcase remarkable adaptability and optimization capabilities. Nonetheless, the significant computational resources required for their real-time deployment pose a formidable challenge. DL frameworks, including CNNs and LSTMs, necessitate substantial computational power, which hampers their feasibility in latency-critical mobile networks. Concurrently, Deep Reinforcement Learning (DRL) models, particularly within multi-agent configurations, demand extensive resources for training, complicating their real-time implementation. As we advance towards 6G, future research endeavors should pivot towards the development of lightweight models and the refinement of efficient training methodologies such as model compression \citep{luan2023channelformer}, quantization, transfer learning and knowledge distillation \citep{odeyomi2022review,zhang2022learning}. Moreover, the adoption of edge computing to decentralize computational tasks could alleviate these computational burdens, facilitating the real-time deployment of these advanced AI techniques in AISM \citep{loven2019edgeai}, thus bridging the existing gap and fostering significant advancements in the field.

\subsection{Federated Learning (FL) for Collaborative AISM}
The rise of FL in AISM (12.90\%) marks a promising path towards addressing privacy concerns and facilitating standardized, collaborative AI/ML-based spectrum management solutions. FL's decentralized machine learning approach allows for models to be trained locally on devices, with only model updates being shared, which is ideal for applications requiring data privacy and security. The integration of deep learning with FL (Deep FL) and the combination of FL with reinforcement learning principles (FRL) illustrate the adaptability of FL to specific AISM needs. However, FL faces challenges such as data and device heterogeneity, communication overheads, and ensuring secure model aggregation. Addressing these challenges requires research into efficient communication protocols, secure aggregation techniques, and strategies for handling, Not Independently and Identically Distributed (non-IID) data across devices [S58]. Standardizing FL-based solutions could facilitate widespread adoption, ensuring compatibility and interoperability across various devices and networks, thus enhancing spectrum management efficiency while adhering to privacy and security requirements.

\subsection{Enhancing transparency in AISM}
RQ1 (Section \ref{section_RQ1}) revealed that B5G studies have utilized AI approaches such as Deep Learning (DL), Machine Learning (ML), Federated Learning (FL), and Reinforcement Learning (RL) techniques for resource, channel, and beam management. However, these approaches exhibit a "black box" nature, posing challenges in understanding their decision-making processes.
For instance, the decentralized nature of Federated Learning complicates the discernment of individual data's impact on the global model, while the intricate architectures of ML obscure the underlying logic guiding decision-making. Similarly, the trial-and-error nature of RL presents challenges in elucidating the rationale behind specific actions. Notably, deep learning methods, employed by 46\% of AISM approaches, inherently exhibit complexity and interpretability issues \citep{wang2021applications}. However, none of the reviewed studies have adequately addressed this concern, indicating a critical area for future research exploration.
To tackle these challenges effectively, a concerted effort towards enhancing the transparency and efficiency of DL models is imperative. Techniques such as explainable AI and model interpretability frameworks offer promising avenues for shedding light on the decision-making processes of DL models, thereby ensuring their trustworthiness in critical applications \citep{fiandrino2022toward, khan2023explainable}. This focus on interpretability not only enhances the reliability of AISM solutions but also fosters greater understanding and acceptance within the broader telecommunications community. 
    \subsection{Integrating Large Language Models (LLMs)} The integration of LLMs into AISM presents a novel research direction with the potential to revolutionize how spectrum resources are allocated, managed, and optimized. While our SLR has highlighted the prevalent use of deep learning techniques in AISM, the application of state-of-the-art LLMs remains unexplored. 
    Given their demonstrated success in various downstream tasks, LLMs could offer unprecedented advancements in AISM by leveraging their superior natural language processing capabilities to interpret, predict, and automate decision-making processes related to spectrum management \citep{jiang2023large,chen2023big,bariah2023large}. The integration of LLMs promises a revolution in spectrum management, enabling the predictive and adaptive allocation of frequency bands through enhanced analytics. This approach could significantly optimize network functions such as localization, beam-forming, and spectrum management in novel network environments \citep{bariah2023large}.
    To fully harness the capabilities of LLMs in AISM, a multi-faceted approach is essential. This encompasses tailoring LLMs with specialized training on real-world spectrum data and refining model architectures to align with the intricate dynamics of spectrum management. Additionally, assessing the computational efficiency and interpretability of these models within existing network infrastructures is critical to ensure their practical applicability and performance. Throughout this process, strict adherence to ethical guidelines and regulatory mandates is paramount to uphold privacy, fairness, and transparency. Therefore, we urge the research community to delve into these areas, paving the way for advanced spectrum management solutions in next-generation networks.
     
\subsection{Need for Real-World Datasets}
The success of deep learning and AI/ML models in real-world spectrum management heavily depends on the quality and diversity of the underlying datasets. Our SLR reveals a significant reliance on simulated and synthetic datasets, with over 65\%  of studies in AI/ML-based spectrum management using such data. This reliance raises questions about these models' ability to generalize to the real world's unpredictability and complexity.
Although invaluable for their realism and direct applicability, real-world datasets are utilized in a smaller fraction of studies. These datasets often come from controlled environments with limited variability, which may not fully capture the dynamics of real-world spectrum usage. The disparity in dataset usage underscores a critical gap in current research practices: the need for more comprehensive and varied real-world datasets.
This scenario underscores a pivotal challenge for the field: enhancing the robustness and effectiveness of AI/ML models through access to diverse, large-scale, real-world datasets. Addressing this challenge is crucial for the advancement of spectrum management technologies, necessitating a concerted effort to collect, share, and utilize real-world data that accurately reflects the complexities of modern telecommunications environments.

 \subsection{Imbalance in AISM S \& P Research}
The exploration of attacks on AISM in RQ2 (Section \ref{sec:securityandprivacy}) reveals a pronounced imbalance in research focus. An analysis of study distribution across various types of attacks—Spoofing (7), Jamming (9), Eavesdropping (18), Evasion (7), Data Leakage (4), Poisoning (4), Inference (2), and Backdoor (2)—highlights a significant skew towards Eavesdropping, with 18 studies dedicated to this threat. This emphasis not only underscores the perceived criticality of Eavesdropping within the spectrum management domain but also points to a comparative under-representation of other attack vectors, particularly inference and backdoor attacks, each with merely 2 studies. This disparity suggests an urgent need for a re-balanced research agenda that extends its focus to under-explored adversarial threats, ensuring a comprehensive understanding and development of defense mechanisms against a wider array of attacks to bolster the security resilience of B5G networks.
Moreover, the distribution of security studies are across different AISM applications helps highlight where more research is needed. While studies focusing on channel estimation and traffic prediction are more common, with 13 and 9 studies, respectively, there's a notable lack of research on how beam tracking might be attacked. This is concerning because beam management is crucial for efficient communication in B5G networks. Additionally, there's little to no research on how data poisoning might affect DSA or how backdoor attacks could impact the process of selecting beams. These gaps suggest that certain vulnerabilities are being overlooked. It's clear that security research in AISM needs to be more evenly spread to cover all areas, including those currently neglected. Taking a comprehensive approach that covers all aspects of spectrum management and potential security threats is essential to protect and strengthen the infrastructure of future B5G networks.

\subsection{Enhancing AI/ML-based S \& P  in B5G Networks}
Given the ongoing developments in 3GPP Releases, particularly the exploration of AI/ML applications in network resource management, air interface applications, and system support, a critical future direction lies in addressing the unresolved challenges of security and privacy within these AI/ML implementations. The technical reports from Release 17 and 18, such as TR37.817 and TR33.877, highlight the need for robust security mechanisms to safeguard user privacy, prevent data poisoning attacks, and ensure secure information transfer between network nodes.
Our RQ2 (Table \ref{table:RQ2_cybersecurity_defenses}) discussed some defenses proposed in the literature. However, we believe that future efforts should focus on developing standardized solutions for these security challenges, incorporating advanced AI/ML techniques to detect and mitigate threats in real-time and enhance the privacy of users within the 5G and beyond networks. This entails creating a comprehensive framework that not only secures AI/ML model transmissions but also provides a resilient infrastructure against attacks that exploit the dynamic nature of AI/ML-based systems. Collaborative efforts among industry, academia, and standardization bodies are essential to formulate guidelines that adapt to the evolving threat landscape, ensuring the safe and trustworthy deployment of AI/ML technologies in critical communication infrastructures.

\subsection{Testbeds for Benchmark and Security Analysis}
Our SLR identified a notable absence of standardized testbeds for assessing the performance and security of AISM systems across diverse applications. We observed in Section \ref{section_RQ1}, that there exist a variations in datasets, system models, and performance metrics among studies targeting the same AISM system hinder fair comparisons and accurate evaluations. This observation aligns with findings presented by Feriani et al. (2023) \citep{feriani2023cebed}, who introduced CeBed—a testbed for channel estimation. CeBed incorporates diverse datasets, system models, and propagation conditions, along with implementations of advanced and conventional baselines. However, it predominantly focuses on Single Input Single Output (SISO) configurations, overlooking the prevalent MIMO setups in B5G networks. Additionally, the AISM field significantly lacks platforms similar to TextAttack \citep{morris2020textattack} and RobustBench \citep{croce2020robustbench}, used in computer vision and NLP domain that thoroughly test AI models' defenses against adversarial attacks. 
Lastly, unlike domains such as computer vision and NLP, the AISM domain lacks transparent sharing of research artifacts for reproducibility and comparison. Therefore, we assert that future efforts should focus on establishing standardized benchmarks and promoting transparency. 
\subsection{Threats to Validity}
Although we followed the guidelines by Kitchenham and Charters \citep{slrguidelines}, our SLR may encounter common threats. These include: \emph{Search Strategy:} Achieving comprehensive coverage of primary studies poses a significant challenge. To counter this, we utilized five data sources to broaden our search, moreover, our search string was refined iteratively, and was augmented by forward and backward snowballing to capture studies potentially omitted. \emph{Bias in Study Selection and Data Extraction:} Subjectivity in selecting studies and extracting data can introduce bias. We adopted a multi-phase selection process and conducted random checks to verify the application of our inclusion and exclusion criteria, which ensures thoroughness. A standardized data extraction form was employed to minimize data extraction bias, facilitating consistent analysis in addressing our research questions.

\section{Conclusion}
The primary objective of this SLR was to examine and consolidate the literature on AISM, focusing on the learning approaches, datasets, performance metrics, and security and privacy concerns in B5G networks. We achieved this by answering two research questions and following SLR methodology to review, analyse and synthesize 110 primary studies. This SLR not only pinpointed research gaps but also offered suggestions for future investigations. Key findings are summarized as follows.
\begin{enumerate}
   \item AI is primarily applied in three areas of Spectrum Management: Resource, Channel, and Beam management.
\item The most prevalent classification model used in AISM studies is based on Deep and Reinforcement Learning, employing CNN as the network architecture.
\item We identified 18 publicly available simulation tools/datasets across AISM applications.
\item The most commonly reported performance metrics are accuracy and error rate.
\item Evasion and Eavesdropping attacks are the most prevalent in the categories of adversarial and inherited attacks, respectively.
\item Defense techniques such as Distillation and Secure Transmission, along with Signal Integrity and anti-jamming measures, are commonly deployed to defend against evasion, eavesdropping, and jamming attacks, respectively.
\end{enumerate}

In conclusion, AISM systems are in their infancy, and many limitations remain. In addition to technical challenges identified in our SLR, standardizing AISM for B5G and beyond faces financial, regulatory, infrastructure, and integration hurdles. Collaboration among telecommunications, regulatory bodies, mobile operators, ML, and cybersecurity experts is crucial to overcome these obstacles. We hope that this SLR will guide these stakeholders by offering research directions and promoting best practices for AISM's evolution in future network generations.
\section{Acknowledgement}
This research paper is conducted under the 6G Security Research and Development Project, as led by the Commonwealth Scientific and Industrial Research Organisation (CSIRO) through funding appropriated by the Australian Government’s Department of Home Affairs. This paper does not reflect any Australian Government policy position. For more information regarding this Project, please refer to  \url{https://research.csiro.au/6gsecurity/.}
 
\bibliographystyle{ACM-Reference-Format}
\bibliography{references}
\end{document}